\documentclass[onecolumn]{emulateapj}
\slugcomment{{\sc Accepted in ApJ:} March 1, 2014}

\usepackage{natbib}
\usepackage{amsmath}
\usepackage{multirow}
\begin{document}

\newcommand{\pasa}{Publications of the Astronomical Society of Australia} 


\newcommand{\HaloRMS}{$155\,\rm\mu Jy\,beam^{-1}$}
\newcommand{\HaloBmaj}{60.7}
\newcommand{\HaloBmin}{48.9}
\newcommand{\HaloBPA}{37.4}

\newcommand{\HaloAtcaRMS}{$28.4\,\rm\mu Jy\,beam^{-1}$}
\newcommand{\HaloAtcaBmaj}{38.4}
\newcommand{\HaloAtcaBmin}{24.7}
\newcommand{\HaloAtcaBPA}{-2.9}

\newcommand{\Ssixten}{$S_{610}=(29\pm 3)\,\rm mJy$} 
\newcommand{\SsixtenNW}{$S_{610}=(19\pm 2)\,\rm mJy$} 
\newcommand{\SsixtenE}{$S_{610}=(1.2\pm 0.2)\,\rm mJy$} 
\newcommand{\SsixtenSE}{$S_{610}=(3.0\pm 0.3)\,\rm mJy$}
\newcommand{\Lonepfour}{$\log (L_{1.4}/\rm W\,Hz^{-1})=25.66\pm 0.12$} 
\newcommand{\LonepfourNW}{$\log (L_{1.4}/\rm W\,Hz^{-1})=25.49\pm 0.11$}
\newcommand{\LonepfourE}{$\log (L_{1.4}/\rm W\,Hz^{-1})=24.32\pm 0.17$} 
\newcommand{\LonepfourSE}{$\log (L_{1.4}/\rm W\,Hz^{-1})=24.65\pm 0.13$}

\newcommand{\NOISEsixten}{$\sigma_{610}=26\,\rm\mu Jy\,beam^{-1}$}
\newcommand{\LX}{$L_{\rm X}=(2.19\pm 0.11)\times 10^{45} \,\rm erg\,s^{-1}$}
\newcommand{\Mfivehund}{$M_{500c}= (1.17\pm 0.17)\times 10^{15} \,M_{\odot}$}

\newcommand{\AlphaTwoNW}{$\alpha^{2.6}_{1.6} = 2.0\pm 0.2$} 
\newcommand{\AlphaTwoE}{$\alpha^{2.6}_{1.6} = 1.2\pm 0.5$} 
\newcommand{\AlphaTwoSE}{$\alpha^{2.6}_{1.6} = 1.2\pm 0.4$} 
\newcommand{\AlphaNW}{$\alpha^{2.1}_{0.6} = 1.19\pm 0.09$}  
\newcommand{\AlphaE}{$\alpha^{2.1}_{0.6} = 0.9\pm 0.1$} 
\newcommand{\AlphaSE}{$\alpha^{2.1}_{0.6} = 1.4\pm 0.1$}
\newcommand{\haloalpha}{$\alpha = 1.2\pm 0.1$}

\newcommand{\AWidthNW}{$34^{\prime\prime}$}
\newcommand{\PWidthNW}{$0.27\,\rm Mpc$}
\newcommand{\ALengthNW}{$1.2^{\prime}$}
\newcommand{\PLengthNW}{$0.56\,\rm Mpc$}
\newcommand{\ALengthE}{$71^{\prime\prime}$}
\newcommand{\PLengthE}{$0.27\,\rm Mpc$}
\newcommand{\unresolvedWidth}{$d_{\rm shock}\leq 23\,\rm kpc$}
\newcommand{\rThreeSigmaHalo}{$r_{\rm H}\simeq 1.1\,\rm Mpc$}
\newcommand{\ShockLimit}{$d_{\rm shock}\leq 23\,\rm kpc$}

\newcommand{\MeanRM}{$\left<\phi\right> = 11\,\rm rad\,m^{-2}$} 
\newcommand{\SigmaRM}{$\sigma_{\phi} = 6\,\rm rad\,m^{-2}$} 
\newcommand{\Bparallel}{$B_{\parallel}\sim 0.01\,\rm \mu G$}
\newcommand{\BCMB}{$B_{\rm CMB} = 11.2\,\rm\mu G$}
\newcommand{\Beq}{$B_{eq}= 0.39\pm 0.09\,\rm\mu G$}  
\newcommand{\IZero}{$2.8\times 10^{-4}\,\rm mJy\,arcsec^{-2}$}
\newcommand{\depth}{$2.2\,\rm Mpc$}

\newcommand{\machNumber}{$\mathcal{M}= 2.5^{+0.7}_{-0.3}$}  
\newcommand{\collisionAngle}{$\phi\geq 11.6\,\rm deg$}  
\newcommand{\ShockSpeed}{$2500^{+400}_{-300} \,\rm km\,s^{-1}$}  
\newcommand{\downStreamSpeed}{$900\pm 100\,\rm km\,s^{-1}$}

\title{THE RADIO RELICS AND HALO OF EL GORDO, 
       A MASSIVE $z=0.870$ CLUSTER MERGER}

\author{Robert R. Lindner\altaffilmark{1,2,$\dagger$}, 
        Andrew J. Baker\altaffilmark{1},  
        John P. Hughes\altaffilmark{1},
        Nick Battaglia\altaffilmark{3},
        Neeraj Gupta\altaffilmark{4,5},
        Kenda Knowles\altaffilmark{6},
        Tobias A. Marriage\altaffilmark{7},
        Felipe Menanteau\altaffilmark{8,9},        
        Kavilan Moodley\altaffilmark{6},
        Erik D. Reese\altaffilmark{10,11}, and
        Raghunathan Srianand\altaffilmark{5}}
                 
\altaffiltext{1}{Rutgers, The State University of New Jersey, 
                 136 Frelinghuysen Road, Piscataway, NJ 
                 08854-8019, USA}
\altaffiltext{2}{Present address: Department of Astronomy, 
                 University of Wisconsin, 
                 Madison, WI 53706, USA}
\altaffiltext{3}{McWilliams Center for Cosmology, Wean Hall, 
                 Carnegie Mellon University, 5000 Forbes Ave., 
                 Pittsburgh PA 15213, USA}
\altaffiltext{4}{ASTRON, the Netherlands Institute for Radio 
                 Astronomy, Postbus 2, 7990 AA, Dwingeloo, The 
                 Netherlands}
\altaffiltext{5}{IUCAA, Ganeshkhind, Pune 411007, India}
\altaffiltext{6}{Astrophysics and Cosmology Research Unit, 
                 University of KwaZulu-Natal, Durban 4041, 
                 South Africa}
\altaffiltext{7}{Department of Physics and Astronomy, Johns
                 Hopkins University, 3400 North Charles 
                 Street, Baltimore, MD 21218-2686, USA}
\altaffiltext{8}{National Center for Supercomputing Applications,
                 University of Illinois at Urbana-Champaign,
                 1205 W. Clark St., Urbana, IL 61801, USA}
\altaffiltext{9}{Department of Astronomy, University of Illinois at
                 Urbana-Champaign, W. Green Street, Urbana, IL 61801, USA}
\altaffiltext{10}{Department of Physics and Astronomy, University of 
                 Pennsylvania, 209 South 33rd St., Philadelphia, 
                 Pennsylvania 19104, USA}
\altaffiltext{11}{Department of Physics, Astronomy, and Engineering, 
                  Moorpark College, 7075 Campus Rd., Moorpark, CA 93021, USA}
\altaffiltext{$\dagger$}{rlindner@astro.wisc.edu}

\begin{abstract}
We present $610\,\rm MHz$ and $2.1\,\rm GHz$ imaging of the massive 
Sunyaev Zel'dovich Effect (SZE)-selected $z=0.870$ cluster 
merger ACT-CL\,J0102-4915 (``El Gordo''), obtained with the 
Giant Metre-wave Radio Telescope (GMRT) and the Australia Telescope Compact 
Array (ATCA), respectively.  We detect two complexes of radio relics
separated by $3.4^{\prime}$ ($1.6\,\rm Mpc$) along the 
system's northwest-to-southeast collision axis that have
high integrated polarization fractions ($33\%$) and 
steep spectral indices ($\alpha$ between 
1--2; $S_{\nu}\propto \nu^{-\alpha}$), 
consistent with creation via Fermi acceleration by 
shocks in the intracluster medium triggered by the 
cluster collision.  From the spectral index of 
the relics, we compute a Mach number \machNumber{} and 
shock speed of \ShockSpeed{}.  With our wide-bandwidth,
full-polarization ATCA data, we compute the 
Faraday depth $\phi$ across the northwest relic and find
a range of values spanning $\Delta \phi = 30\,\rm rad\,m^{-2}$, 
with a mean value of \MeanRM{} and standard deviation \SigmaRM{}.  
With the integrated line-of-sight gas density
derived from new {\em Chandra} X-ray observations, our Faraday depth 
measurement implies \Bparallel{} in the cluster outskirts.  
The extremely narrow shock widths in
the relics (\unresolvedWidth{}), caused by the short 
synchrotron cooling timescale of relativistic electrons at $z=0.870$, 
prevent us from placing a meaningful constraint on the magnetic field
strength $B$ 
using cooling time arguments.
In addition to the relics, we 
detect a large (\rThreeSigmaHalo{} radius), 
powerful (\Lonepfour{}) radio halo with a shape 
similar to El Gordo's ``Bullet''-like X-ray morphology.
The spatially-resolved spectral-index map
of the halo shows the synchrotron spectrum is
flattest near the relics, along the system's 
collision axis, and in regions of high 
$T_{\rm gas}$, all locations associated with recent energy
injection.  The spatial and spectral correlation between
the halo emission and cluster X-ray properties 
supports primary-electron processes like
turbulent reacceleration as the halo production mechanism.
The halo's integrated $610\,\rm MHz$ to $2.1\,\rm GHz$ 
spectral index is a relatively flat
\haloalpha{}, consistent with the cluster's high
$T_{\rm gas}$ in view of previously established 
global scaling relations.
El Gordo is the highest-redshift cluster known to host
a radio halo and/or radio relics, and provides new 
constraints on the non-thermal physics in clusters
at $z>0.6$.
\end{abstract}


\section{Introduction}
Galaxy clusters grow out of peaks in the primordial matter 
distribution of the early Universe and gain mass by 
accreting gas from their environments and by merging with 
other clusters and groups of galaxies.  
Because the physical properties and dynamical state 
of the intracluster medium (ICM) are affected by its 
merger history,  observations of the ICM can be used to 
study the growth and evolution of clusters.
Cluster mergers can inject large amounts of gravitational
potential energy ($\sim 10^{64}\,\rm erg$) 
into the ICM,
and observations of cluster-scale radio synchrotron emission, 
particularly in merging and dynamically disturbed 
systems \citep[e.g., ][]{buote2001,pfrommer2008,brun09,cass11},
indicate that some fraction 
of this energy is directed into accelerating cosmic ray 
electrons to 
ultra-relativistic ($\gamma \sim 10^3$--$10^4$) energies.  
The resulting non-thermal radio 
emission is seen as a radio halo, i.e., diffuse axisymmetric 
emission centered on the cluster, or  as radio relics, 
narrow, extended, filamentary structures located 
near the cluster outskirts (for reviews, see \citealt{ferr08} 
and \citealt{fere12}). 

Competing explanations for nonthermal relic and halo emission
fall in two categories.  ``Primary'' models rely on
acceleration of electrons by shocks (e.g., first-order
Fermi process: \citealt{enss98}) or turbulence
(e.g., turbulent reacceleration process: 
\citealt{brun01}) in the ICM.  
``Secondary'' models posit that cosmic-ray
electrons are produced by inelastic hadronic collisions
between thermal protons and cosmic ray protons, the
latter accelerated by cluster mergers and long lived
\citep{dennison1980}.  Because these models make different
predictions about whether and how the strength and spectral
shape of radio emission relate to the local state
of the ICM, by studying the 
spectral, morphological, and polarization 
properties of radio halos and relics, we can probe the
poorly understood non-thermal properties of
galaxy clusters such as cosmic ray acceleration and
magnetic field profiles.
The geometry of relic systems can also be
used to constrain the collision parameters
of cluster mergers
\citep{vanw11m11}.

One challenge in using relics and halos to probe 
the energy content and magnetic field properties 
of clusters is their relative rarity.  
Only $\simeq 30\%$ of X-ray luminous clusters host 
halos \citep{venturi2007,venturi2008}. Their presence is 
correlated with cluster mass and dynamical state, with the 
most massive and most 
dynamically disturbed clusters showing the highest 
frequencies of halos and 
relics \citep{Cassano2013,Sommer2013}.  
Large radio relics, arcs with lengths $\ge 1\,\rm Mpc$, are found
 only in massive, merging clusters, and even rarer 
double-relic systems are found only in 
binary cluster mergers occurring in the plane of the sky.

The paucity of massive high-redshift clusters combined with 
the steep observed spectral 
index\footnote{flux density $S_{\nu} \propto \nu^{-\alpha}$} 
$\alpha$ of the non-thermal 
emission 
\citep[$\alpha \simeq 1.5$; ][]{ferr08} 
make the study of halos and relics at high redshifts 
especially difficult.  Additionally, energy losses of the 
relativistic electrons to cosmic microwave background (CMB) 
photons through inverse Compton scattering are expected 
to reduce the radiative lifetimes, and therefore 
detectability, of synchrotron-emitting regions.  
Because of the above limitations, catalogs of known radio 
halo and relic clusters only extend to low 
redshifts; the highest-redshift
halo cluster discovered to date is 
PLCK\,G147.3--16.6 \citep[$z=0.65$; ][]{vanweeren2014}.
\citet{nuza12} predict that many more $z>0.3$ 
relics should exist than are currently catalogued, 
with the deficit likely due to lack of observations.
To understand the nature of
non-thermal emission in clusters
throughout cosmic time, we need to increase the
sample of known relics and halos 
at high redshift.

In this work, we present new $610\,\rm MHz$ 
and $2.1\,\rm GHz$ 
observations of the extremely 
massive, $z=0.870$ cluster
ACT-CL\,J0102-4915 
\citep[also known as ``El Gordo,'' and hereafter referred
to by that name; ][]{mena12}, which reveal an associated
radio halo and double radio relics.
As the highest redshift radio-halo
cluster now known, El Gordo can help 
fill in the gaps in our knowledge about
non-thermal cluster physics 
at high redshift when the Universe was only 
half its current age.  
Section \ref{s-elgordo} describes El Gordo,
Section \ref{s-obs} describes our observations and data 
reduction algorithms, Sections \ref{s-relics} and 
\ref{s-halo} present our analyses of and results for the
relics and halo, respectively, and Section 
\ref{s-conclusion} presents our conclusions.
In our calculations, we assume a nine-year $WMAP$ 
cosmology with 
$H_0=70\,\rm km\,s^{-1}\,Mpc^{-1}$, $\Omega_M=0.27$, and 
$\Omega_\lambda=0.73$ \citep{hinshaw2013}. At $z=0.870$, 
$D_{\rm L}=5656\,\rm Mpc$, $D_{\rm A}=1618\,\rm Mpc$, and 
$1\rm\,arcmin$ corresponds to $0.47\rm \,Mpc$.

\section{ACT-CL\,J0102-4915, ``El Gordo''}
\label{s-elgordo}
El Gordo was discovered through its $148\,\rm GHz$ 
Sunyaev-Zel'dovich effect \citep[SZE; ][]{suny72}
decrement by the Atacama Cosmology Telescope 
\citep[ACT; ][]{fowl07} collaboration.
With a SZE centroid 
at $01^{\rm h}02^{\rm m}53^{\rm s}$ 
$-49^{\circ}15^{\prime}19^{\prime\prime}$ (J2000)
and the strongest SZE 
signal in the $455\,\rm deg^{2}$ southern 
ACT survey 
\citep[][]{marr11}, El Gordo was optically
confirmed as a bona fide galaxy cluster at 
$z=0.870$ \citep[][]{mena10} with a  ``bullet''-like merger
morphology revealed in $Chandra$ X-ray imaging \citep{mena12}.
El Gordo's collision axis, identified by the system's
elongation in the X-ray surface brightness and the
relative positions of two optical galaxy density peaks,
lies in the northwest-to-southeast direction at a position
angle of $136^{\circ}$ \citep{mena12}.
Based on the merger morphology,  
\citet{mena12} predict the inclination angle $\phi$ 
(between the collision axis and the plane of the sky)
to be shallow, in the range $\phi=15$--$30^{\circ}$.
The cluster is also the most significant detection in the 
(overlapping) $2500\,\rm deg^2$ survey \citep[][]{will11} of the 
South Pole Telescope \citep[SPT;][]{carl11}.
El Gordo has an observed-frame 
$0.5$--$2.0\,\rm keV$ X-ray luminosity of 
\LX{}
and an integrated X-ray temperature of 
$k_{\rm B}T_{\rm gas}=14.5\pm 0.1\,\rm keV$ \citep{mena12}.
By combining X-ray, SZE, and
velocity dispersion measurements, \citet{mena12} estimate
a total mass\footnote{
The mass estimate assumes the conversion 
factor of $f=M_{200a}/M_{500c}=1.85$ from \citet{mena12}, 
who adopt the same cosmological parameters as this work.}
within $r_{500c}$, the radius containing 500 times
the critical density of the universe,  
of \Mfivehund{}.
This mass estimate is compatible with subsequent
analyses of strong lensing by \citet{Zitrin2013} and
of weak lensing by \citet{jee2013}.

\section{Observations and data reduction}
\label{s-obs}

\subsection{2.1\,GHz ATCA} 
We used the Australia Telescope Compact 
Array (ATCA) to acquire 2.1GHz imaging of
El Gordo.   The Compact Array Broadband Backend 
\citep[CABB; ][]{wils11} has
2048 $1\,\rm MHz$-wide channels that span 
$1.1$--$3.1\,\rm GHz$.  
The observations were obtained in two
installments (see Table \ref{t-obs-elgordo}):  12 hours
in the extended 6A configuration in 
December 2011 (PI: Baker), and 8 hours in the compact 
1.5B configuration in April 2012 
(PI: Lindner).  Calibration used 
the flat-spectrum, radio-loud quasars PKS\,1934-638 (bandpass
and flux calibration) and PKS\,0047-579 (initial phase 
calibration).  Observations of PKS\,1934-638
bracketed each observing session, while 
phase-tracking observations of PKS\,0047-579 
were taken every 30 minutes.
We used the software package 
MIRIAD \citep{saul95} to calibrate, flag, invert, and 
clean the visibility data.

Radio frequency interference (RFI) is significant in the $2.1\,\rm GHz$ 
band at the ATCA.  We removed RFI from the data both manually
using the MIRIAD tasks \texttt{pgflag} and \texttt{blflag}, and
automatically using the MIRIAD task \texttt{mirflag} 
\citep{midd06}.
Baseline 1--2 ($uv$ distance $=194\,\lambda$) in the 2012 observations 
contained powerful broad-spectrum RFI and was 
entirely flagged.
In total, 22\% of visibilities were flagged.
We used the task \texttt{invert} to produce
multi-frequency synthesis (MFS) continuum images and
\texttt{mfclean} to remove artifacts caused by
the dirty beam and produce cleaned images.  The data were
then self-calibrated, first allowing the phase to vary, then
phase and amplitude together.  The final
image was made with 
\texttt{robust}\footnote{The \texttt{robust} parameter
controls how the data are weighted in the $uv$ plane
prior to inversion.  The weighting converges to 
``uniform'' weighting for \texttt{robust}$=-2$ and to
``natural'' weighting for \texttt{robust}$=+2$.}$=0$ $uv$ 
weighting, giving a synthesized beam 
with dimensions 
$6.1^{\prime\prime}\times 3.1^{\prime\prime}$ and
a position angle ${\rm PA}=-1.9^{\circ}$. 
The image RMS noise at the 
phase center is 
$\sigma_{2.1}=8.2\,\rm\mu Jy\,beam^{-1}$ for an
effective frequency of $\nu=2.15\,\rm GHz$.
The $2.1\,\rm GHz$ map of El Gordo is shown in
Figure \ref{f-2100}.

The 2011 and 2012 observations have parallactic angle 
coverages  of $120^{\circ}$ and $230^{\circ}$, respectively, 
allowing for polarization calibration.
\citet{schn11} note that the instrumental 
polarization leakage across the
CABB bandpass varies with frequency.  Therefore, 
during gain calibration, we solved for 
leakage corrections separately
in each of eight 
$256\,\rm MHz$-wide subbands using
the \texttt{gpcal} option \texttt{nfbin}$=8$.

\subsection{610\,MHz GMRT}
We used the 30-antenna Giant Metre-wave Radio Telescope (GMRT)  
to acquire $610\,\rm MHz$ imaging of El Gordo.  Observations 
were carried out in August 2012 (PI: Lindner) in three four-hour
tracks. We used the quasar 3C48 for bandpass
and flux calibration and J\,0024-420 for initial phase 
calibration.
The data have 256 130\,kHz-wide channels
for a total bandwidth of $33\,\rm MHz$
and an effective frequency of $607.7\,\rm MHz$.
The $uv$ data were
calibrated and imaged using the Common 
Astronomy Software Applications (CASA) package 
(Version 4.0.0)\footnote{\texttt{http://casa.nrao.edu/}}.
The tracks were visually inspected to remove powerful
RFI spikes that were constant in 
either time or frequency, and 
transient RFI signals were removed using the automated
flagging algorithm AOFLAGGER \citep{offr10, offr12}.

\begin{figure}
    \centering
    \includegraphics[scale=0.8]{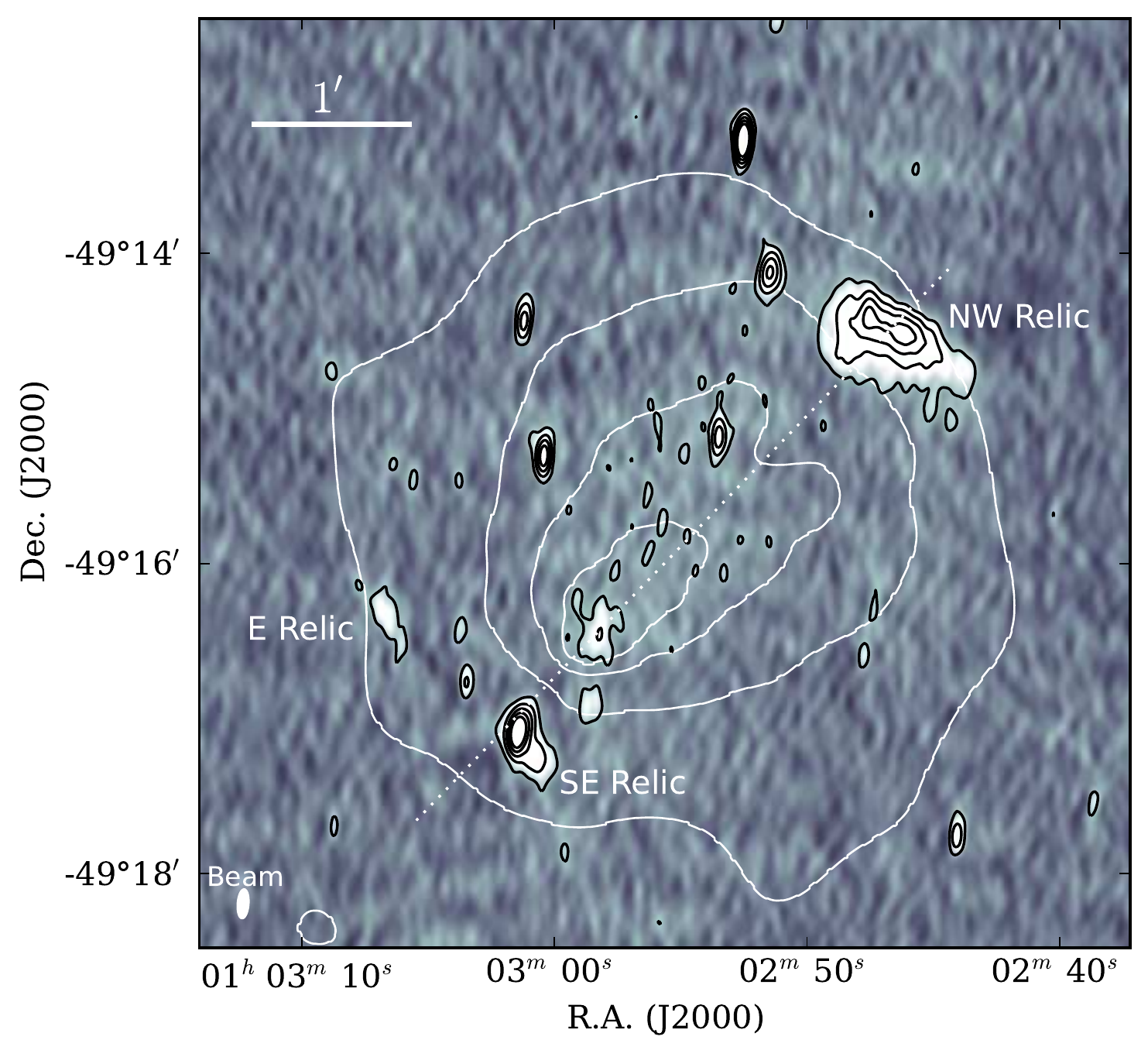}
    \caption{$610\,\rm MHz$ GMRT image of El Gordo.  The color 
        stretch is $[-10\sigma, 10\sigma]$ and black contours 
        are shown at $5,15,30,50,75\sigma$,
        where $\sigma=26\,\rm\mu Jy\,beam^{-1}$.
        White contours represent logarithmic $0.5-2.0\,\rm keV$ 
        X-ray surface brightness \citep{mena12}.
        Labels mark the locations of the NW, E, and SE relics.          
        The dotted line represents the estimated collision axis
        ($\rm P.A.=136^{\circ}$) from \citet{mena12}.
        The image includes all GMRT data and is produced with 
        multi-frequency synthesis and multi-scale clean.
        The synthesized beam is shown in the lower left corner.}
    \label{f-610}
\end{figure}

The GMRT's non-coplanar $uv$ dataset was imaged using
three facets and 512 $w$-projection planes.
The final self-calibrated map is shown in 
Figure \ref{f-610} and has an RMS
sensitivity of \NOISEsixten{} 
and a synthesized beam of 
$11.0^{\prime\prime}\times 4.0^{\prime\prime}$ at 
position angle $-4.3^{\circ}$.  The elongated
synthesized beam is due to the low maximum
elevation ($\simeq 20^{\circ}$) when 
El Gordo is observed from the GMRT site.

Our measured flux densities at $610\,\rm MHz$ and 
$2100\,\rm MHz$ are affected by both statistical image 
noise and systematic errors in the bootstrapped 
absolute flux densities of the flux calibrators.
We have included this extra systematic error, 
10\% for GMRT \citep{baars1977,chandra2004} and 5\% for 
ATCA \citep{reynolds1994}, in the quoted uncertainties 
for all measured flux densities and derived quantities.

\section{Radio relics}
\label{s-relics}

\citet{mena12} identified two 
$843\,\rm MHz$ sources on opposite sides of El Gordo
in archival data from the Sydney University Molonglo 
Sky Survey \citep[SUMSS;][]{mauc03} that were
aligned with the collision axis and coincided with  
the locations of possibly shocked thermal 
gas, as traced by an unsharp-masked image of the 
$Chandra$ $0.5$--$2.0\,\rm keV$ X-ray map \citep{mena12}.  
This morphology is the signature arrangement of 
well-studied double radio relics in binary major 
cluster mergers at lower 
redshifts \citep[e.g., ][]{vanw11m4}.
In this arrangement, the filamentary relics
are spatially extended perpendicular
to the system's collision axis, identifying an
outward-propagating shock front in the ICM.
The outermost edges (with respect to the cluster 
center) of the radio relics correspond 
to the leading edges of the shock fronts. The 
trailing emission traces downstream shocked 
regions, which can be identified by their 
steeper spectral indices compared to the 
leading edges, caused by synchrotron 
spectral aging.  The flattest spectral
indices are thus found in the leading edges of the relics
where recent energy injection has taken place.
Our $2.1\,\rm GHz$ ATCA and $610\,\rm MHz$ GMRT maps 
confirm the presence of the relics in El Gordo, whose 
properties we detail below.

\subsection{Geometries}
\label{ss-geometries}
\begin{figure}
    \centering
    \includegraphics[scale=0.8]{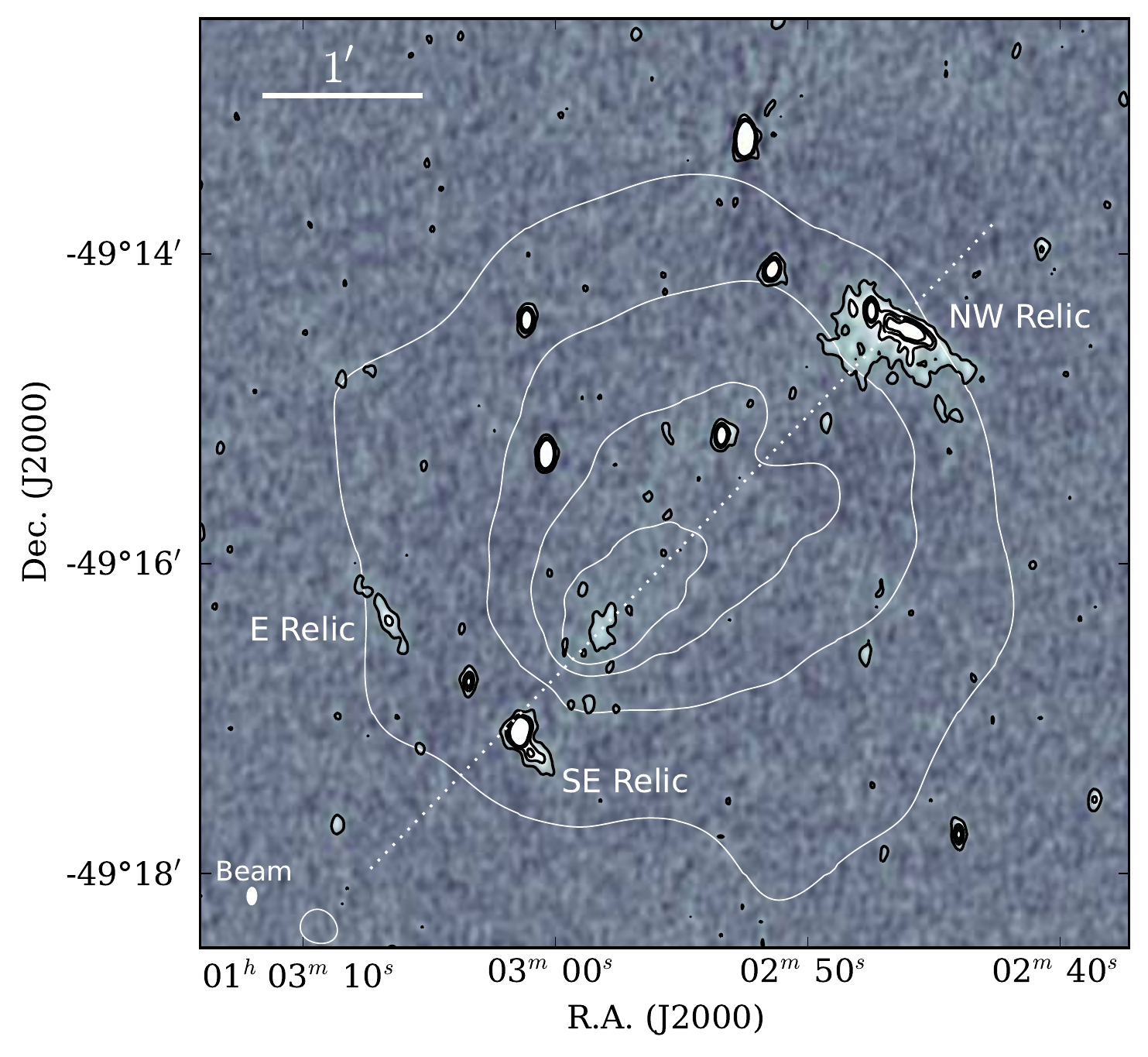}
    \caption{$2.1\,\rm GHz$ ATCA image of El Gordo.  The color stretch is
        $[-10\sigma, 10\sigma]$ and black contours are shown at $3,10,15,20\sigma$,
        where $\sigma=8.2\,\rm\mu Jy\,beam^{-1}$.
        White contours represent logarithmic $0.5-2.0\,\rm keV$ 
        X-ray surface brightness \citep{mena12}.
        Labels mark the locations of the NW, E, and SE relics.   
        The dotted line represents the estimated collision axis
        ($\rm P.A.=136^{\circ}$) from \citet{mena12}.
        The synthesized beam is shown in the lower left corner. The image
        includes all ATCA data and is produced with multi-frequency synthesis 
        and multi-scale clean.}
    \label{f-2100}
\end{figure}

The elongation directions of all relics are
perpendicular to the collision axis 
(see Figures \ref{f-2100} and \ref{f-610}), suggesting
the relics are created by shock waves in the ICM
of the cluster merger \citep{enss98}.
Our high-resolution imaging reveals the 
northwest SUMSS source to be an extended 
radio relic (hereafter
referred to as the NW relic), which is resolved 
in both length and width by our observations.
The NW relic has a total length of \ALengthNW{} 
(\PLengthNW{}), a thickness of
\AWidthNW{} (\PWidthNW{}), and a bright, unresolved
ridge of emission on its northwest (outer) edge.
The internal structure of the NW relic is complex; 
it contains a trailing filament with an unresolved
width that
is extended 
along a position angle
offset from
the NW relic's leading edge.
The flux from the southeast SUMSS source is due to
a compact radio source (C10)  with
$S_{2.1\,\rm GHz}={(1.9\pm 0.1)\,\rm mJy}$ 
(see Table \ref{t-sources}),
probably unrelated to the cluster, 
superposed on a
much fainter extended component that 
we interpret as a likely radio relic (hereafter referred
to as the SE relic).
A third component of extended filamentary emission
located $\sim 1^{\prime}$ northeast 
of the SE relic we will refer to as 
the E relic; it has a resolved length of
\ALengthE{} (\PLengthE{}).
The widths of the E and SE relics and 
the width of the bright
leading edge of the NW relic are all
unresolved by our observations,
indicating very narrow shocked 
regions of \ShockLimit{} (Figure \ref{f-2100}).

\begin{figure}
    \centering
    \includegraphics[scale=0.8]{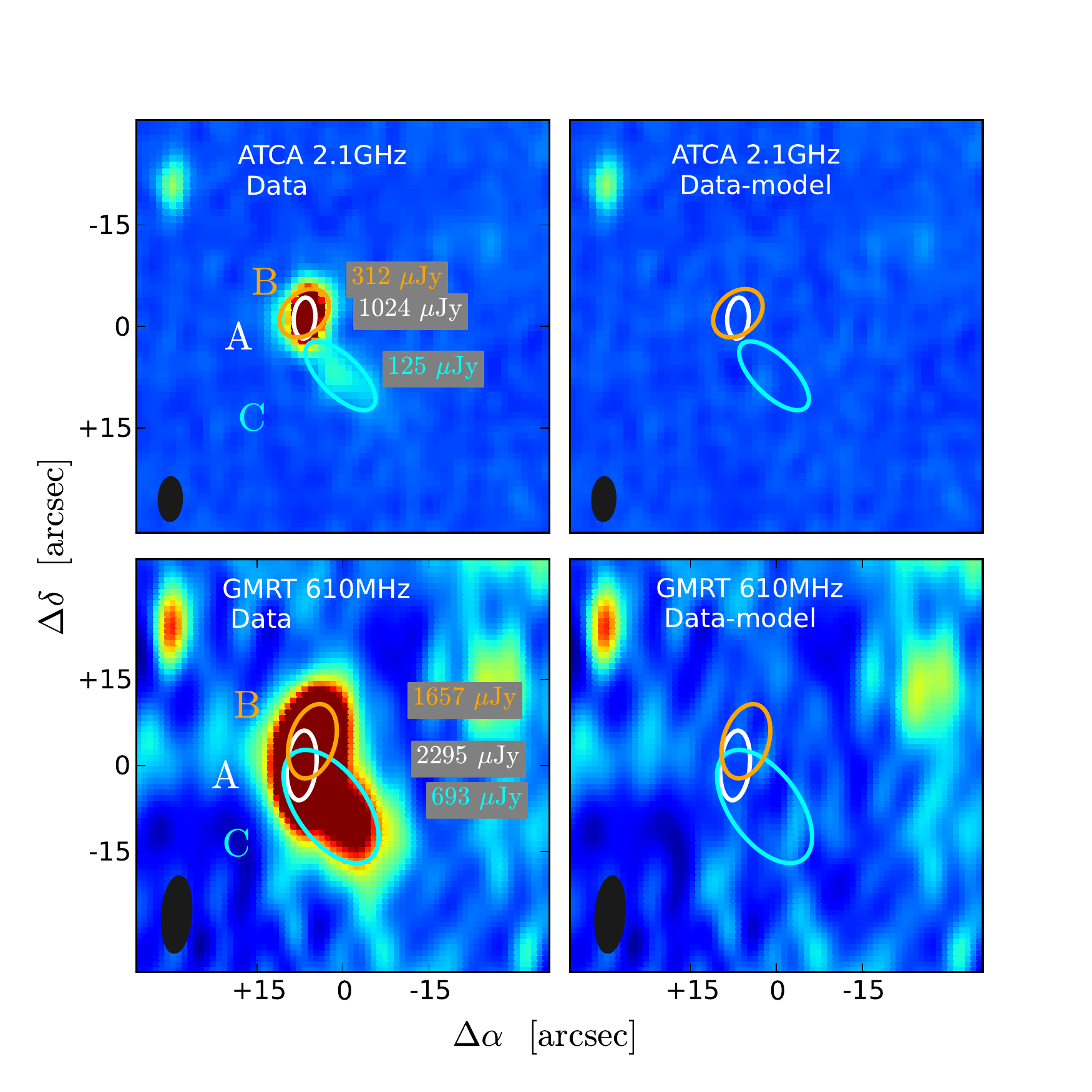}
    \caption{Gaussian decomposition of the SE relic.  
    The emission is modeled with three elliptical 
    Gaussian components. The contours represent the 
    FWHMs of the best-fit elliptical Gaussians at
    $2.1\,\rm GHz$ (above) and $610\,\rm MHz$ (below).
    The original images ({\em left}) and model-subtracted
    residual images ({\em right}) are shown.
    The best-fit peak flux densities for the components 
    are shown in the original image panels.
    The white, orange, and cyan contours (and text) represent 
    components ``A'', ``B'', and ``C'', respectively 
    (see Section \ref{ss-geometries}).  The centroids,
    shape parameters, and numbers of components of the 
    elliptical Gaussian fits were not forced to match
    between $2.1\,\rm GHz$ and $610\,\rm MHz$;  the agreement
    between frequencies is support for the 
    reality of the SE relic (component C).}
    \label{f-components}
\end{figure}

To study the photometric properties of the SE relic, 
we subtract C10 from the image data by modelling 
its emission as a collection of 
elliptical Gaussian sources.  We find that three Gaussian
components are sufficient to produce residuals consistent
with noise.  Figure \ref{f-components} displays the
best-fit model configuration, which consists of a 
dominant source whose shape is consistent with the
synthesized beam (``A''), a moderate flux density, resolved 
component closely associated with the point source (``B''), 
and a faint very extended component elongated perpendicular
to El Gordo's collision axis (``C'').  Our subsequent 
photometric analysis of the relics is performed in 
images with components ``A'' and ``B'' subtracted out.
Table \ref{t-properties} lists the integrated 
photometric properties of the relics.

\begin{figure}
    \centering
    \includegraphics{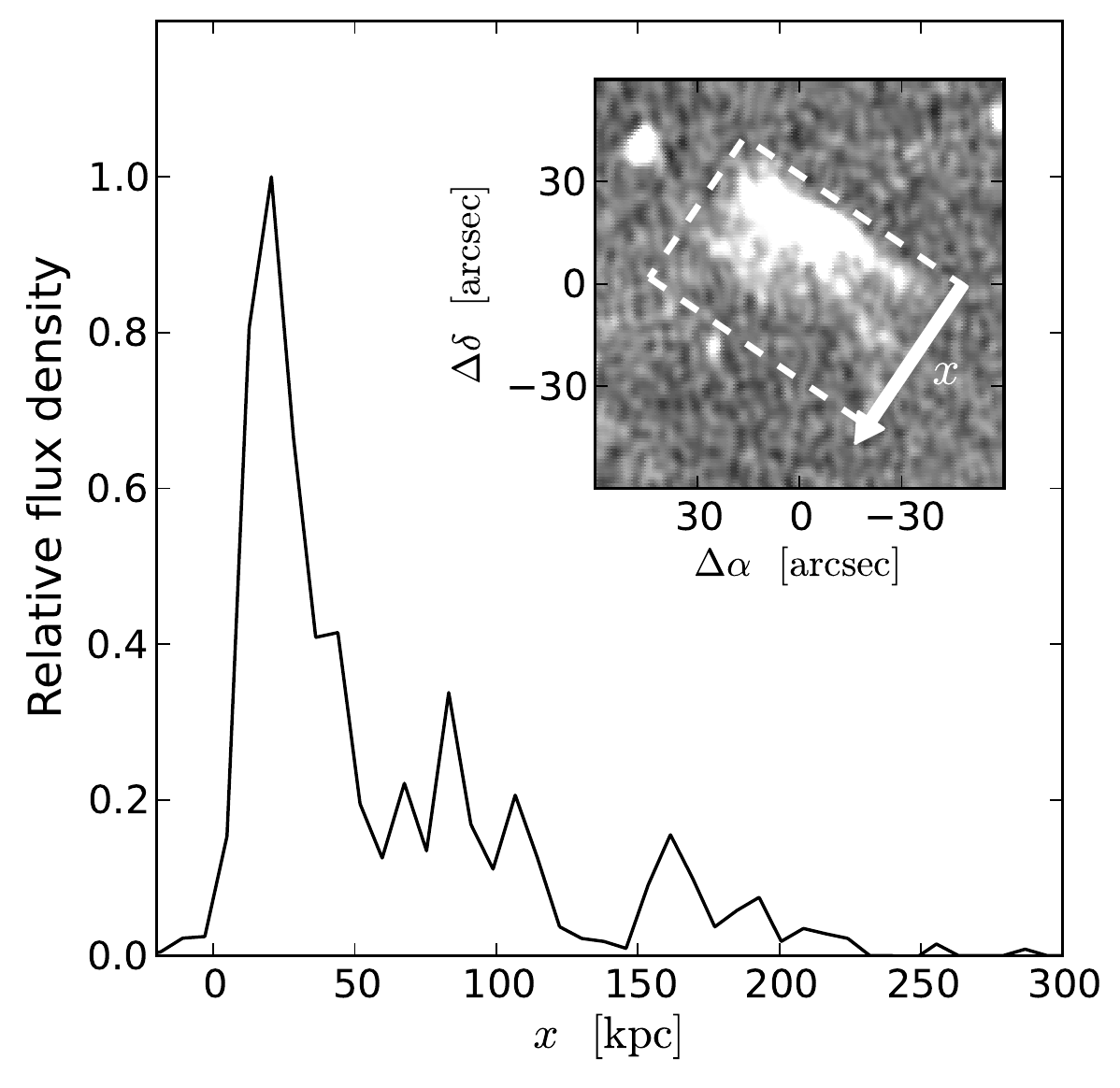}
    \caption{Projected radial profile of NW relic.
        The projection region is a 
        $1.3^{\prime}\times 0.8^{\prime}$ box
        at P.A.=$55^{\circ}$, shown in the inset.
        The bright leading edge drops quickly, and an
        extended tail of emission continues out to
        $\sim 200\,\rm kpc$.  The extended tail is due to
        projection and to the superposition of multiple
        fainter filaments.}
    \label{f-nw-profile}
\end{figure}

Figure \ref{f-nw-profile} presents a profile of
the NW relic projected perpendicular to the collision
axis, showing the leading unresolved edge
followed by an extended trail of low-level 
signal likely caused by the spherical projection
of a 3D shock structure and filamentary 
substructures in the shock material.  Similar 
extended tails are seen in simulations
of cluster mergers by \citet{vanw11m11}.

\subsection{Spectral indices}

Our multi-wavelength data allow us to produce
a $610\,\rm MHz$/$2.1\,\rm GHz$ spectral index
($\alpha^{2.1}_{0.6}$) 
map using the full 
GMRT and ATCA images, and 
a $1.6\,\rm GHz$/$2.6\,\rm GHz$ spectral 
index ($\alpha^{2.6}_{1.6}$) map 
using the upper and lower halves
(i.e., $1.1-2.1\,\rm GHz$ and $2.1-3.1 \,\rm GHz$) 
of the ATCA bandpass.
We need not worry about incomplete recovery
of large-scale emission in the relics 
because even the largest dimension of the 
NW relic ($1.3^{\prime}$)
is smaller than the angular scale 
on which we expect
our ATCA data to begin resolving out emission
($\gtrsim 1.9^{\prime}$) given the 
minimum $uv$ distance ($1.8\,\rm k\lambda$) at
the high-frequency end of the bandpass.
The GMRT data extend to a lower $uv$ distance
and preserve emission on scales even larger 
than the cluster.
Each pair of images was 
smoothed to a common circular beam
($11^{\prime\prime}$ for $\alpha_{0.6}^{2.1}$
and $8^{\prime\prime}$ for $\alpha^{2.6}_{1.6}$) and 
then 
clipped at a $4\sigma$
level before
we computed the spectral index\footnote{The two-point spectral index is defined by 
$\alpha \equiv -\log\left(S_{\nu_1}/S_{\nu_2}\right)/\log\left(\nu_1/\nu_2\right)$,
with uncertainty
$\sigma_{\alpha}=\frac{ \sqrt{(\sigma_{S_1}/S_1)^2+(\sigma_{S_2}/S_2)^2}}{\ln\left(\nu_1/\nu_2\right)}$.} 
$\alpha$.

\begin{figure}
    \centering
    \includegraphics[scale=0.8]{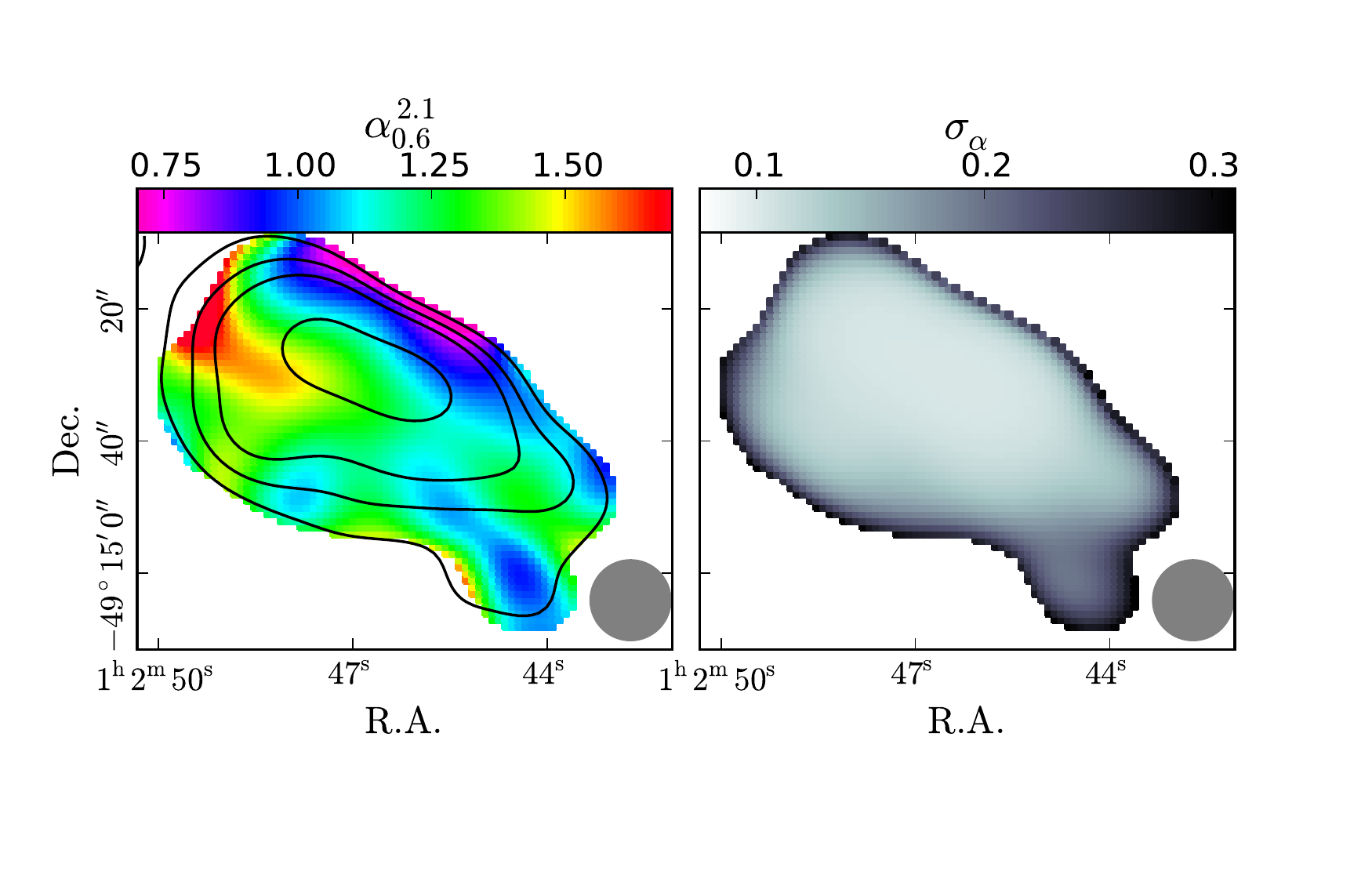}
    \caption{Spectral index ($\alpha^{2.1}_{0.6}$) 
        map of the NW relic (left) and corresponding
        uncertainty (right).  
        Contours=[0.3,0.6,1.0,3.0]\,$\rm mJy\,beam^{-1}$.
        The filled grey circle represents the effective 
        resolution ($12^{\prime\prime}$).}
    \label{f-relics-nw-610}
\end{figure}

\begin{figure}
    \centering
    \includegraphics[scale=0.8]{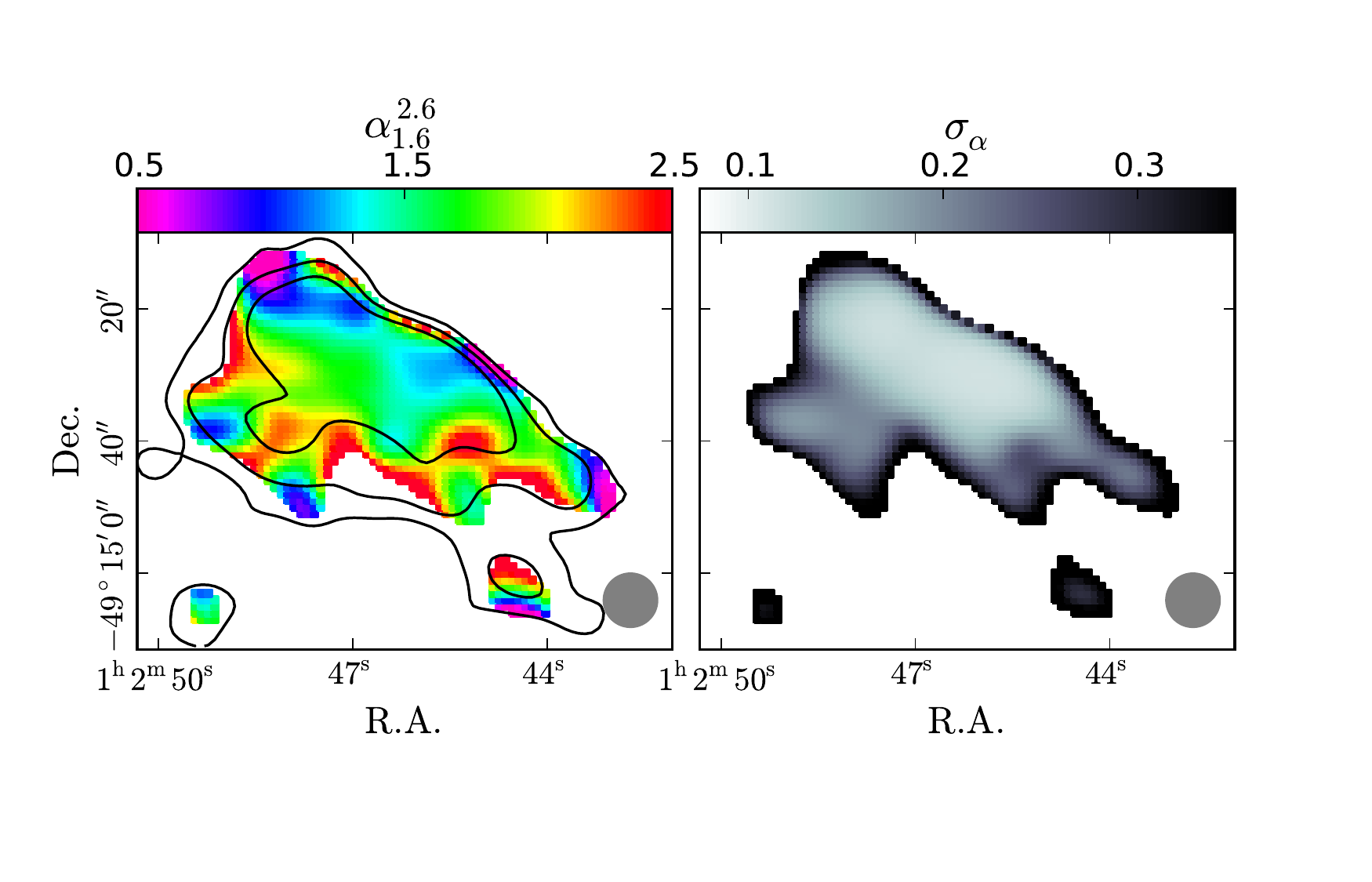}
    \caption{Spectral index ($\alpha^{2.6}_{1.6}$) 
        map of the NW relic (left) and corresponding
        uncertainty (right).  
        Contours represent $2.1\,\rm GHz$ intensity
        levels of 30, 100, and $200\,\rm \mu Jy\,beam^{-1}$.
        The grey circle represents the effective resolution 
        ($8^{\prime\prime}$).}
    \label{f-relics-nw-2100}
\end{figure}

\begin{figure}
    \centering
    \includegraphics[scale=0.8]{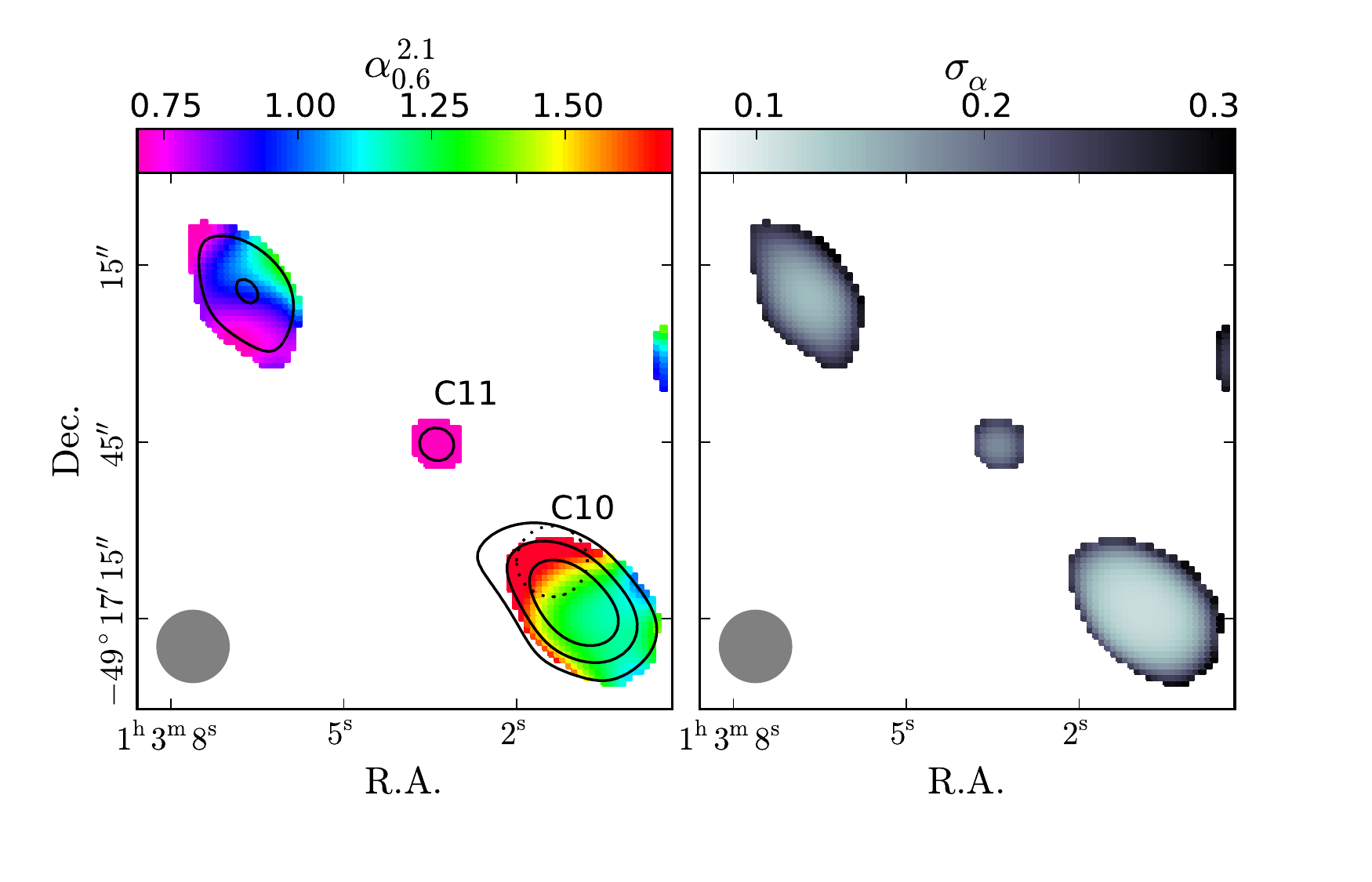}
    \caption{Spectral index ($\alpha^{2.1}_{0.6}$) 
        map of the E and SE relics (left) and corresponding 
        uncertainty (right).  
        Contours represent $610\,\rm MHz$ intensity
        levels of $[0.3,0.6,1.0,3.0]\,\rm mJy\,beam^{-1}$.
        The dotted circle marks the position of subtracted
        compact source C10.
        The grey circle represents the effective resolution
        ($12^{\prime\prime}$).}
    \label{f-relics-se-610}
\end{figure}

\begin{figure}
    \centering
    \includegraphics[scale=0.8]{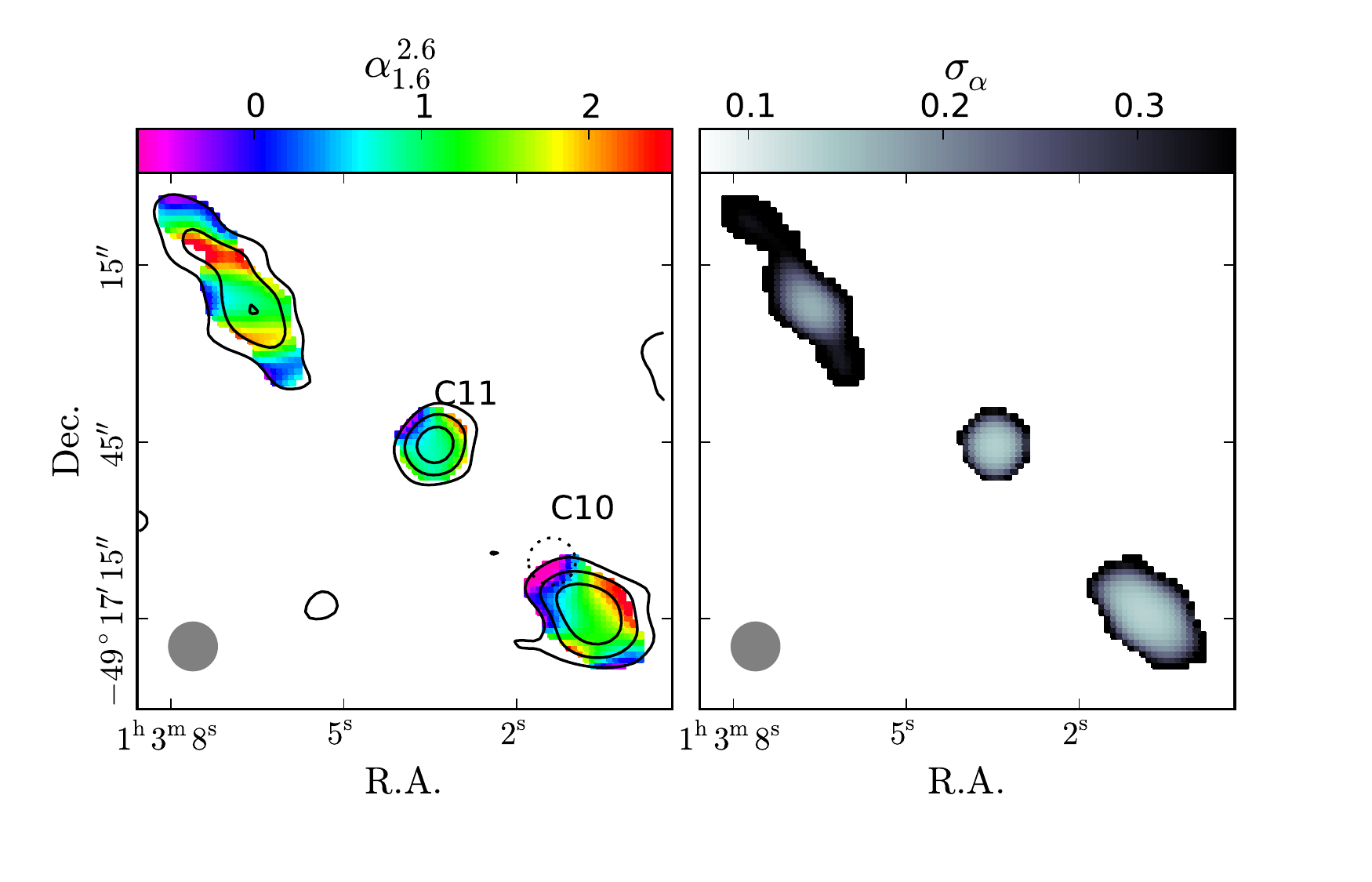}
    \caption{Spectral index ($\alpha^{2.6}_{1.6}$) 
        map of the E and SE relics (left) and corresponding
        uncertainty (right).
        Contours represent $2.1\,\rm GHz$ intensity
        levels of 30, 100, and $200\,\rm \mu Jy\,beam^{-1}$.
        The dotted circle marks the position of subtracted
        compact source C10.
        The grey circle represents the effective resolution 
        ($8^{\prime\prime}$).}
    \label{f-relics-se-2100}
\end{figure}

Figures \ref{f-relics-nw-610} and \ref{f-relics-nw-2100}
present the $\alpha^{2.1}_{0.6}$ and $\alpha^{2.6}_{1.6}$
spectral index maps of the NW relic, respectively.
The integrated spectral 
indices 
are \AlphaNW{} and \AlphaTwoNW{}.
Figures \ref{f-relics-se-610} and \ref{f-relics-se-2100} show
$\alpha^{2.1}_{0.6}$ and $\alpha^{2.6}_{1.6}$ for 
the the E and SE relics, respectively.  The E relic has 
integrated spectral index values of 
\AlphaE{} and \AlphaTwoE{}, and the SE relic
spectral indices are \AlphaSE{} and \AlphaTwoSE{}.

For the NW and E relics, where unlike the SE relic there
are no uncertainties in the flux due to 
contamination by a nearby radio source,
$\alpha^{2.6}_{1.6}$ is steeper than
$\alpha^{2.1}_{0.6}$ due to
spectral aging effects at high frequencies.
The E relic resembles the leading edge of
the NW relic in its narrow width and flatter 
spectral index, which are likely due to 
decreased projection effects and recent energy
injection.

\subsection{Rotation measure and $B_{\parallel}$}
\label{ss-rm}
Magnetized plasma along the line of sight
to polarized emission will cause Faraday rotation
of the polarization angle of the emitted photons
by an amount
$\Delta \Psi\;[\rm rad] = {\rm RM}\times \lambda^2$, where
$\Psi$ is the polarization angle 
($\Psi=0.5\tan^{-1}(U/Q)$), $\lambda\;[\rm m]$ is the
wavelength of observation, and $\rm RM\;[rad\, m^{-2}]$ 
is the rotation
measure.
Our wide-bandwidth, full-polarization ATCA
data allow us to measure the RM of the relic signal and 
constrain the integrated product of the 
parallel component of the magnetic field $B_{\parallel}$ and 
the free electron density $n_e$ through the cluster 
outskirts at a projected cluster-centric distance 
of $871 \,\rm kpc$.  For a single source of polarized emission, 
the rotation measure is equal to the 
Faraday depth
${\rm \phi}\,[{\rm rad\,m^{-2}}]=0.81\,\int_{\rm source}^{\rm observer} 
n_e B_{\parallel} dl$ \citep{burn66}, 
with $B_{\parallel}$ in $\rm \mu G$, 
$n_e$ in $\rm cm^{-3}$, and $dl$ in $\rm pc$.
To avoid ``$n\pi$'' errors and to allow for the detection of
multiple RM components, we compute the Faraday spectrum, 
the complex polarized surface brightness per unit 
Faraday depth $F(\phi)$ $[\rm Jy\,beam^{-1}\,\phi^{-1}]$,
using the RM synthesis algorithm \citep{bren05}
as implemented by the Astronomical Image Processing 
System (AIPS\footnote{\texttt{http://www.aips.nrao.edu/}}) in
the task \texttt{FARS}.

\begin{figure}
    \centering
    \includegraphics{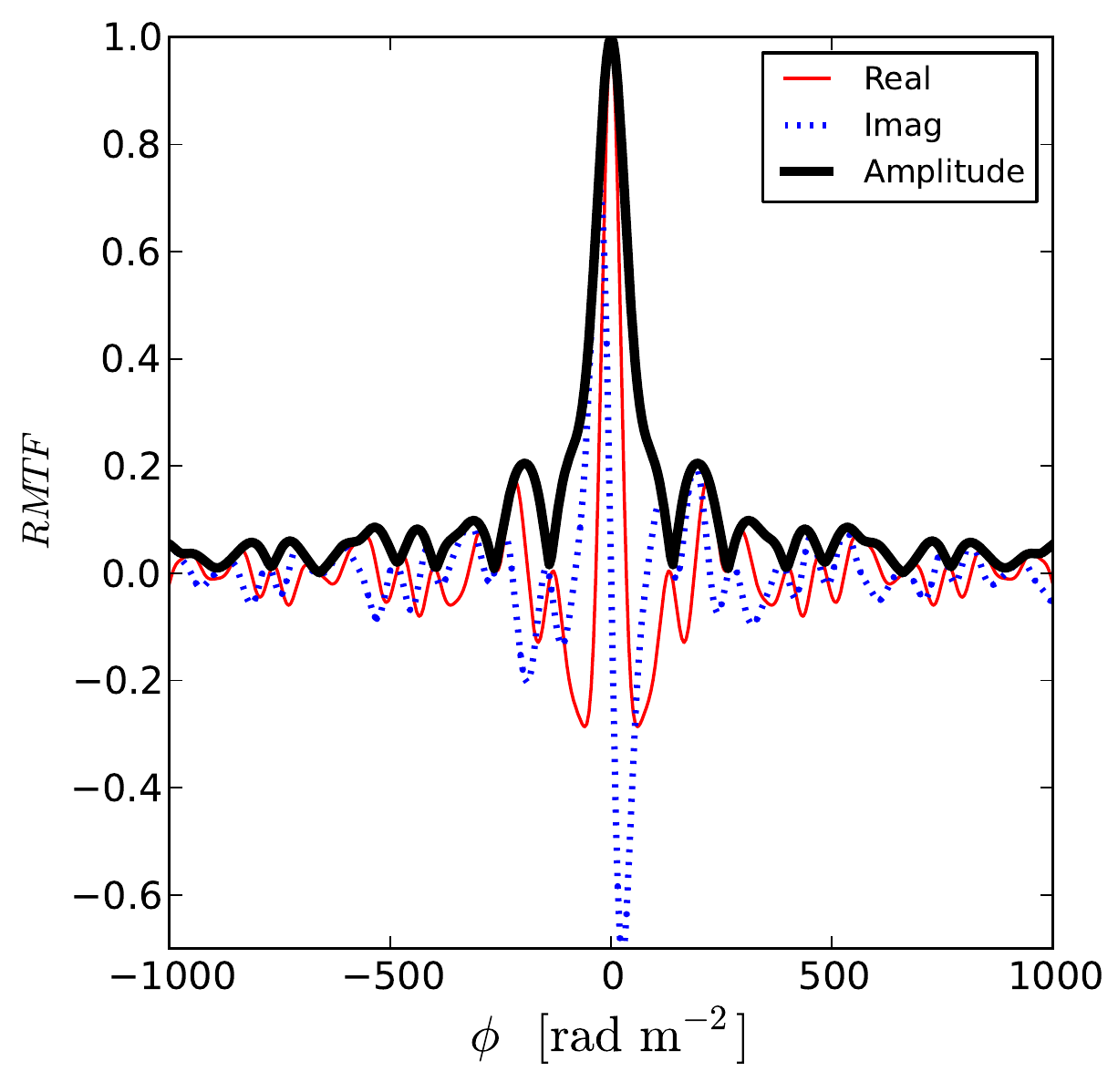}
    \caption{Rotation measure transfer function of ATCA $2.1\,\rm GHz$
        observations.
        The red and blue lines represent the real and imaginary
        components; the black line represents the total amplitude.}
    \label{f-rmtf}
\end{figure}

We first produced $10\,\rm MHz$-wide $Q$ and $U$
images with $3^{\prime\prime}$ pixels, which
were then individually corrected for the
varying primary
beam attenuation across the wide CABB bandwidth, 
smoothed to a common resolution
with a circular synthesized beam of 
FWHM $11^{\prime\prime}$,
and assembled
into a single data cube before
we ran \texttt{FARS}.
The region of maximum sensitivity is 
a circle with radius 
$\sim 14^{\prime}$, entirely covering the cluster, which 
is set by the high-frequency end of the 
bandpass.
The $10\,\rm MHz$-wide channel maps were then 
weighted by the 
inverse variance of the noise within a central
$8^{\prime}$ box.  The effective 
$\left<\lambda^2\right>$ 
of the data is $\left<\lambda^2\right>=0.20\,\rm m^{2}$.
The FWHM of the RM transfer function (RMTF), 
shown in Figure \ref{f-rmtf},
is $\delta_{\rm RM}=145\,\rm rad\,m^{-2}$.

\begin{figure}
    \centering
    \includegraphics{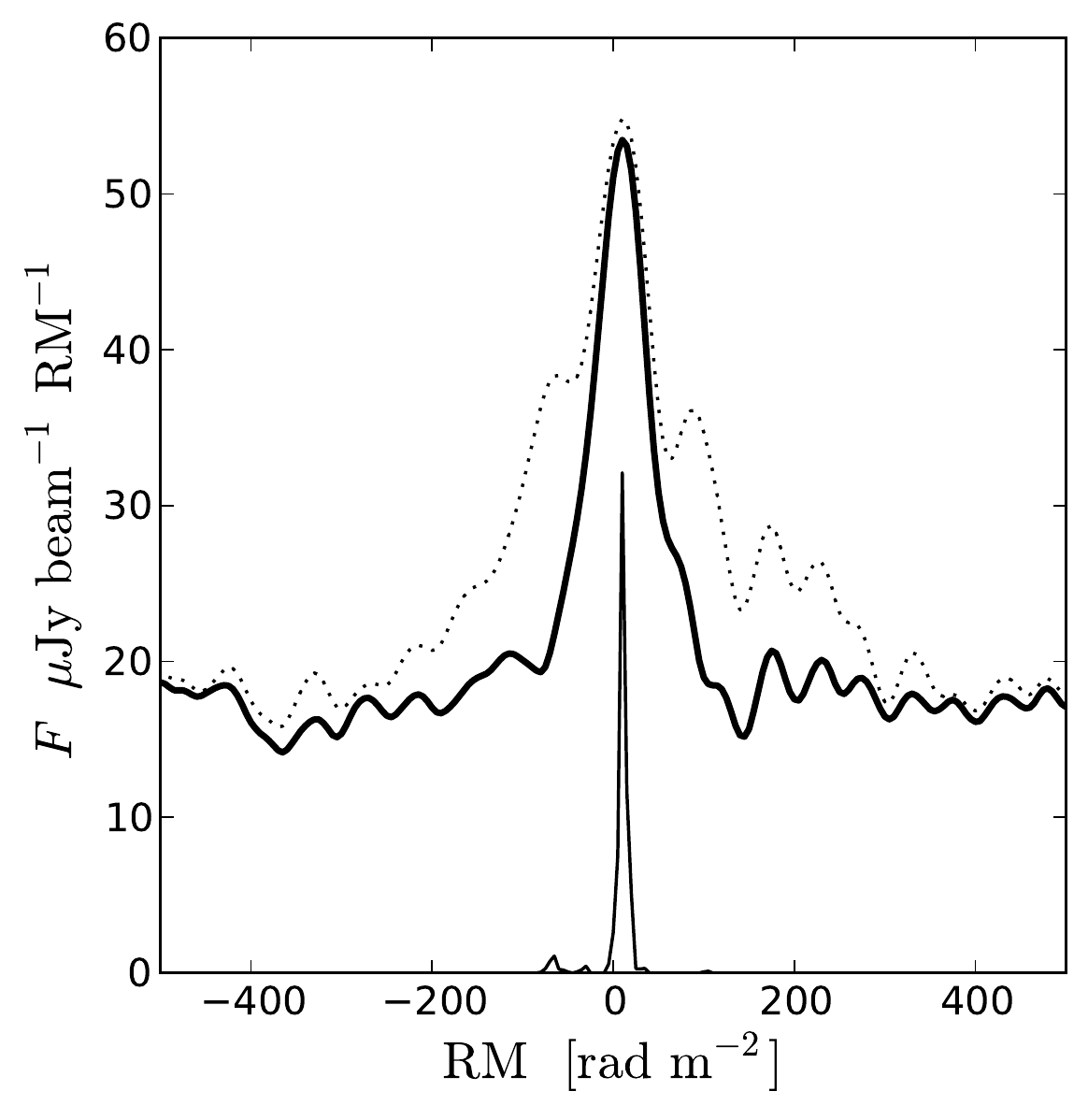}
    \caption{Integrated RM synthesis spectrum for the
        NW relic using ATCA ($2.1\,\rm GHz$) data.  
        The dotted and thick solid lines show the
        uncleaned and cleaned spectra, respectively.  The
        thin solid line below represents the cleaned model
        components (amplitudes have been scaled by $\times 3$
        for visual clarity).}
    \label{f-rm-spectrum}
\end{figure}

\begin{figure}
    \centering
    \includegraphics[scale=0.70]{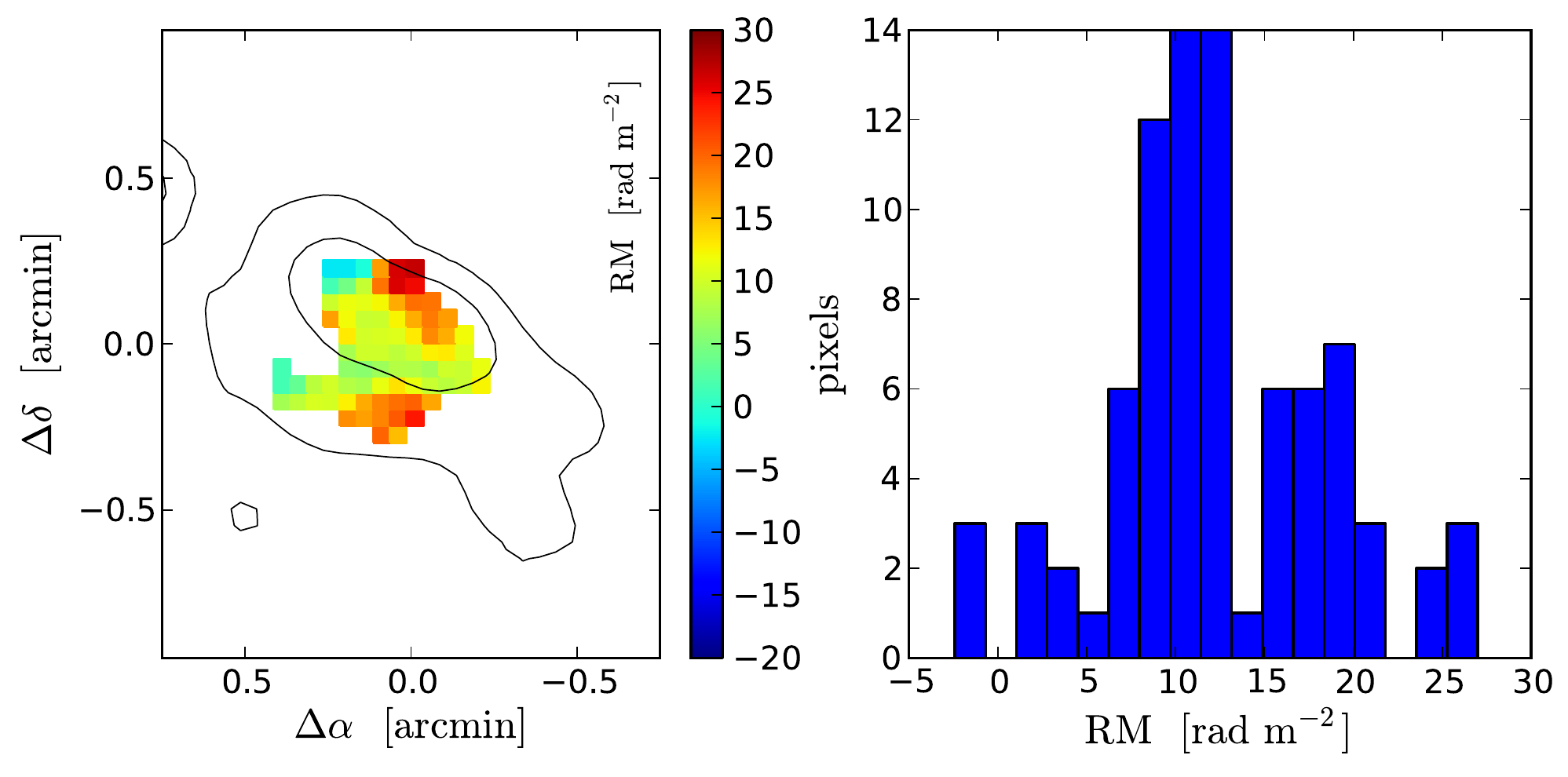}
    \caption{{\em Left:} RM value of dominant
        RM component in each pixel across the NW relic.
        RM values are clipped at $>3\sigma$ in
        the polarized signal image.  Stokes $I$ contours
        are shown at 30 and $300\,\rm\mu Jy\,beam^{-1}$.
        {\em Right:} Distribution of  
        RM values in the left panel.
        The uncertainty in the RM centroid of a given component
        ranges between 5--$10\,\rm rad\,m^{-2}$.  RM analysis
        uses ATCA ($2.1\,\rm GHz$) data.}
    \label{f-rm-image}
\end{figure}

The spatially-integrated Faraday depth spectrum for 
the polarized emission in the NW relic is 
shown in Figure 
\ref{f-rm-spectrum}.  We find that the 
polarized signal in each pixel is 
consistent with a single, dominant Faraday 
component.  Therefore, the polarized emission 
from the NW relic is likely originating from a 
single plane behind the magnetized plasma of the
cluster and foreground material.
We find no other significant
Faraday components out to
$\pm 10^4\,\rm rad\,m^{-2}$.
We next used the AIPS task \texttt{AFARS}
to produce an image showing the Faraday depth of
the dominant component in each pixel across
the relic (Figure \ref{f-rm-image}).
The Faraday depth contribution 
from the Galaxy is estimated using the 
RM maps of \citet{oppe12}.
Within a $2^{\circ}$-radius circle centered 
on El Gordo, the mean Galactic RM
value is $1.0\,\rm rad\,m^{-2}$, 
with a range between $-0.1\,\rm rad\,m^{-2}$
and $+1.9\,\rm rad\,m^{-2}$.

The Faraday depths for the NW relic have a 
Galactic-subtracted mean of \MeanRM{} and
a standard deviation of \SigmaRM{}; 
across the relic they span a range from 
$-5\,\rm rad\,m^{-2}$ to $+25\,\rm rad\,m^{-2}$.
The uncertainty in each Faraday depth value is
given by $\delta_{\rm RM}/(2\times {\rm SNR})$, 
where SNR is the signal-to-noise ratio in the
Faraday spectrum. 
For the pixels shown in Figure \ref{f-rm-image}, 
the uncertainties range between 
5--$10\,\rm rad\,m^{-2}$.
The Faraday depths of the NW relic 
are similar to the RM values of the radio 
relics in the Coma cluster presented recently by
\citet{bonafede2013}, who 
find a mean value
$\left<{\rm RM}\right>=14\,\rm rad \,m^{-2}$
with  
$\left<\sigma_{\rm RM}\right>=17\,\rm rad\, m^{-2}$.
For a system which only has a single Faraday component 
we would expect consistency between RM and Faraday depth 
measurements, although we note that RM 
measurements derived using $d\Psi/d\lambda^2$ with 
sparse $\lambda^2$ coverage \citep[e.g., ][]{bonafede2013}
are subject to greater systematic uncertainties than 
Faraday depth measurements derived using RM synthesis.
We find that the range of Faraday depths across the
NW relic significantly 
exceeds that due to the RMTF, likely
reflecting the changing magnetic field strength
and orientation ($\int B_{\parallel}\,dl$) in 
the ICM.  \citet{bonafede2013} also find
significant variations in RM across the Coma
relics and suggest they are due to turbulence
in the ICM.

Recently, \citet{osul12} studied the Faraday
spectra of four bright ($>1\,\rm Jy$) 
radio-loud active galactic nuclei using
the ATCA/CABB at $2.1\,\rm GHz$ and 
obtained RMTF widths 
$\delta_{\rm RM}\simeq 60\,\rm rad\,m^{-2}$, 
i.e., narrower than ours.
The reason for this difference in $\delta_{\rm RM}$ is 
our goal of detecting the $<1\,\rm mJy$
polarized emission from faint relic structure
in El Gordo, which leaves us more susceptible to low-level RFI.
RFI generally increases at low frequencies, 
thereby reducing the weights of the high-$\lambda^2$
coverage and increasing the width of the RMTF, given by
$\delta_{\rm RM}=2\sqrt{3}/(\lambda^2_{\rm max}-\lambda^{2}_{\rm min})$.
We note that when we use 
equal weighting instead of inverse-variance weighting 
across the $\lambda^2$ channels, 
our RMTF becomes significantly narrower with 
$\delta_{\rm RM}=79\,\rm rad\,m^{-2}$, although
the RM centroid uncertainties $\sigma_{\rm RM}$
are not improved in this case
due to the increased noise from including all
low-frequency channels.

The electron column density along the 
line-of-sight to the NW relic is computed by integrating 
the cylindrically-deprojected electron density 
model from \citet{mena12} from the cluster
mid-plane out to infinity, and is 
$\int n_{e}\,dl = 954\,\rm cm^{-3}\,pc$.
We thus estimate the mean value for the parallel
component of the magnetic field as 
$\left<B_{\parallel} \right> = \left<\phi\right>/(0.81 \int n_{e}\,dl)$, 
and find \Bparallel{} in the outskirts of El Gordo.  
Caution should be used when interpreting the above
magnetic field strength due to the significant
assumption (necessary given the lack of
knowledge about the magnetic field's topology) 
that the projected magnetic field strength is 
constant along the line of sight.

\subsection{Polarization}

Highly polarized synchrotron emission
is evidence of a highly aligned magnetic field
in an emitting region.  Such alignment can be caused
by shocks sweeping up ICM material at
the locations of radio relics. The total amplitude 
of linearly polarized emission $P=\sqrt{Q^2+U^2}$,
and the total fractional polarization is
given by $f_{P}=P/I$.  Uncertainties in $P$ and 
$f_{P}$ follow a Rice distribution, which has 
non-zero mean.  In our maps, we account for this 
bias by multiplying the polarized signal $P$ by
the correction factor $f_R=\sqrt{1-{\rm SNR}^{-2}}$
\citep{ward74}, valid for $\rm SNR>1.0$ per pixel.

\begin{figure}
    \centering
    \includegraphics[scale=0.8]{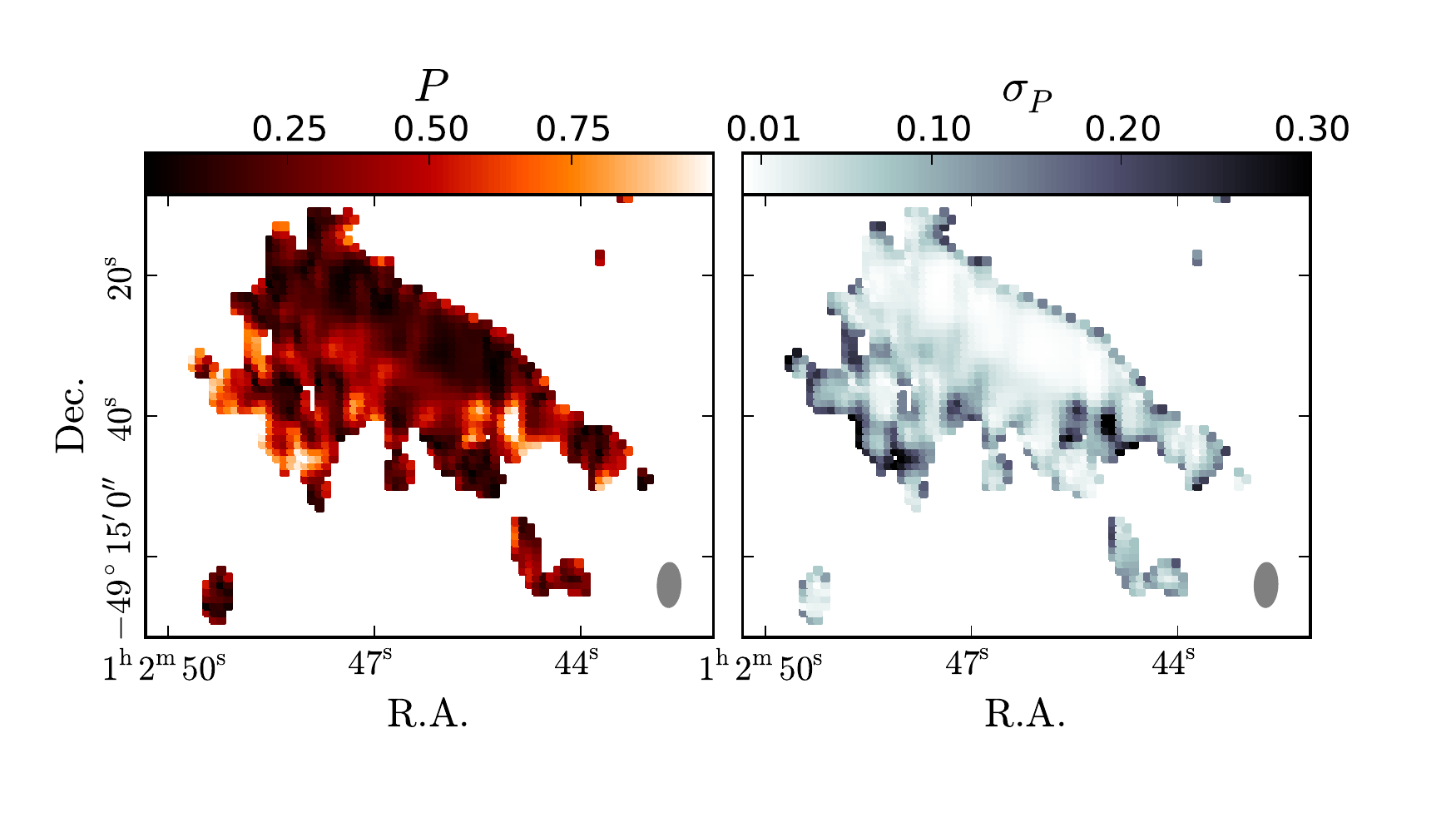}
    \caption{Polarization fraction $f_{P}$
        across the NW relic (left), with
        corresponding uncertainty (right).}
    \label{f-polarization}
\end{figure}

The NW and E relics have similar integrated 
polarization values of $33\pm 1\%$ and $33\pm 3\%$, 
respectively.
Figure \ref{f-polarization} presents the
fractional polarization across the NW relic.
The maximum mean fractional polarization found 
within a beam-sized aperture across the NW relic 
is $67\%$, close to the maximum possible value
for synchrotron emission with $\alpha=1$ (75\%)
or $\alpha=2$ \citep[81\%; ][]{rybicki1979}.
The fractional polarization is reduced in the 
leading edge of the NW relic and enhanced in the 
trailing regions.  This gradient 
is unlikely to be caused by depolarization from 
internal or external Faraday dispersion, which
would instead tend to {\em reduce} the degree of 
polarization in emission with steeper spectral 
indices, and may instead be due to an increased 
alignment of the magnetic fields in the 
post-shock region.  
The high degree of polarization in the relics 
further supports the finding of \citet{mena12},
who argue based on X-ray morphology that the
collision is occurring nearly in the plane 
of the sky, and is also compatible with our 
constraint on the angle between the collision 
axis and the plane of the sky of \collisionAngle{} 
(see Section \ref{ss-shock}).

\begin{figure}
    \centering
    \includegraphics[scale=0.95]{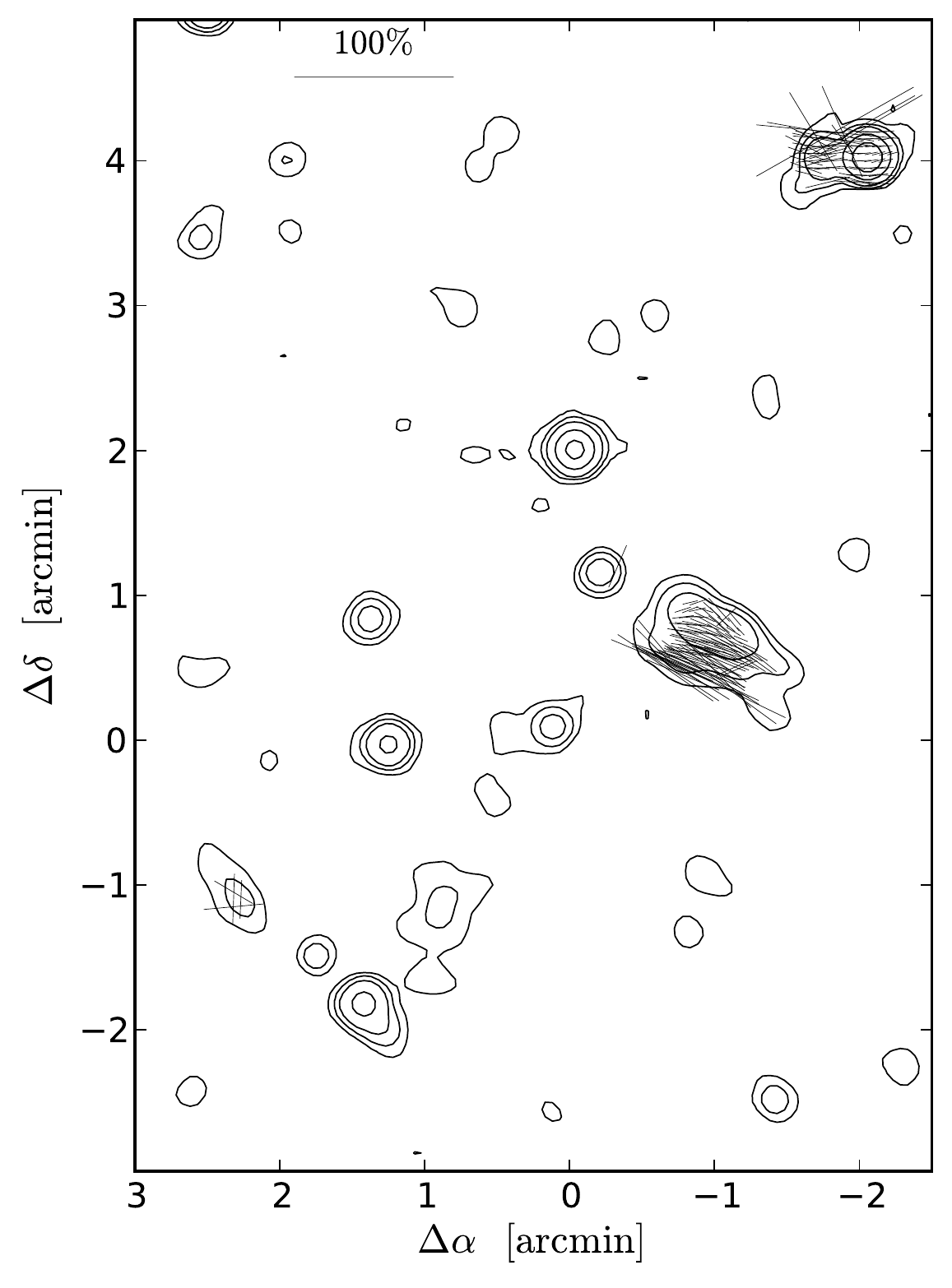}
    \caption{Polarization fraction $f_{P}$ and 
        angle $\Psi$ of relic emission.
        Polarization angles have been corrected for
        rotation measure and further rotated $90^{\circ}$, 
        indicating the direction of $B_{\perp}$.
        The polarized source in the northwest is
        SUMSS\,J010240-491118 with $S_{843}=1.3\,\rm mJy$, 
        and is not associated with the
        cluster. The $f_P=100\%$ 
        scale bar is shown at the top of the panel.
        The
        polarization computation is clipped when the Stokes $I$
        S/N is $<5\sigma$. Contours of Stokes $I$ emission are 
        drawn at $3,10,25,100,$ and $300\sigma$. The 
        RMS noise in the Stokes $I$ image 
        is $\sigma=12\,\rm\mu Jy\,beam^{-1}$.}
    \label{f-rm-vectors}
\end{figure}

We use the RM map to ``derotate'' the  $Q$  and $U$
datasets to produce an image of the intrinsic
polarization angle $\Psi$ across the relic.
Figure \ref{f-rm-vectors} shows the polarization
angles after further rotation by $90^{\circ}$
so that line segments indicate the direction
of the projected magnetic field.  We find that the 
projected magnetic field is aligned with the 
relic's elongation axis in the NW relic, and is 
at least partially aligned in the fainter 
E and SE relics as well.

\subsection{Shock properties}
\label{ss-shock}
We constrain the magnetic field strength $B$
at the location of the NW relic using the shock width and 
spectral index \citep[e.g., ][]{vanw10}. We assume that a thin 
shock with upstream and downstream speeds $v_1$ and $v_2$, 
respectively, propagates along the collision axis. Assuming the 
electrons are energized by the first-order Fermi acceleration 
mechanism \citep{drur83}, the compression ratio $r$ determines
the index of the particle energy distribution function 
$p=(r+2)/(r-1)$, which is related to the spectral index of the 
synchrotron emission via $\alpha=(p-1)/2$. The compression 
ratio is related to the Mach number of a thin shock through the 
Rankine-Hugoniot jump condition: 
\begin{equation}
\frac{1}{r}=\frac{\gamma-1}{\gamma+1}+\frac{2}{\gamma+1}
\frac{1}{\mathcal{M}^2},
\label{e-mach}
\end{equation}
where we assume 
$\gamma\equiv c_P/c_V=5/3$ for a monatomic gas.
For the leading edge (up to an offset of $6^{\prime\prime}$) of the
NW relic, 
$\left< \alpha\right>=0.86\pm 0.15$, giving 
$p=2.7 \pm 0.3$, and $r=2.7^{+0.4}_{-0.3}$.
The estimated
Mach number $\mathcal{M}$ of the shock is found using
Equation \ref{e-mach} to be \machNumber{}.
We measure the gas temperature in a region
that is downstream relative to the NW relic to
be $kT_2 = 11\pm 2 \,\rm keV$, then estimate the upstream
electron temperature using
\begin{equation}
kT_e = kT_2\,\frac{(\gamma+1)-(\gamma-1)r}{(\gamma+1) - (\gamma-1)r^{-1}} = 3.8^{+1.7}_{-1.6}\,\rm keV,
\end{equation}
and compute a sound speed of 
$c_s=\sqrt{\gamma\,kT_{e}/\mu m_e}=1000\pm 200 \,\rm km\,s^{-1}$, 
where we use the molecular weight of 
solar-metallicity plasma, $\mu=0.615$.
Using the compression ratio from above, this 
implies a down-stream velocity 
$v_2=$\downStreamSpeed{} and
shock speed (upstream velocity) 
$v_1=$\ShockSpeed{}.
If we assume the cluster is undergoing 
its first pass of the collision, the shock
speed can be interpreted as an upper limit 
on the collision speed, which should be lower 
than the shock speed due
to the cluster's declining mass density profile
in the regions near the NW relic and to the 
possible infall of material in the upstream 
region.  
Using the $\sim 600\,\rm km\,s^{-1}$ difference in
radial velocity between the two concentrations of
galaxies in the SE and NW \citep{mena12} allows us
to constrain 
the angle between the collision axis and 
the plane of the sky to be \collisionAngle{} (84\% confidence).

\begin{figure}
    \centering
    \includegraphics{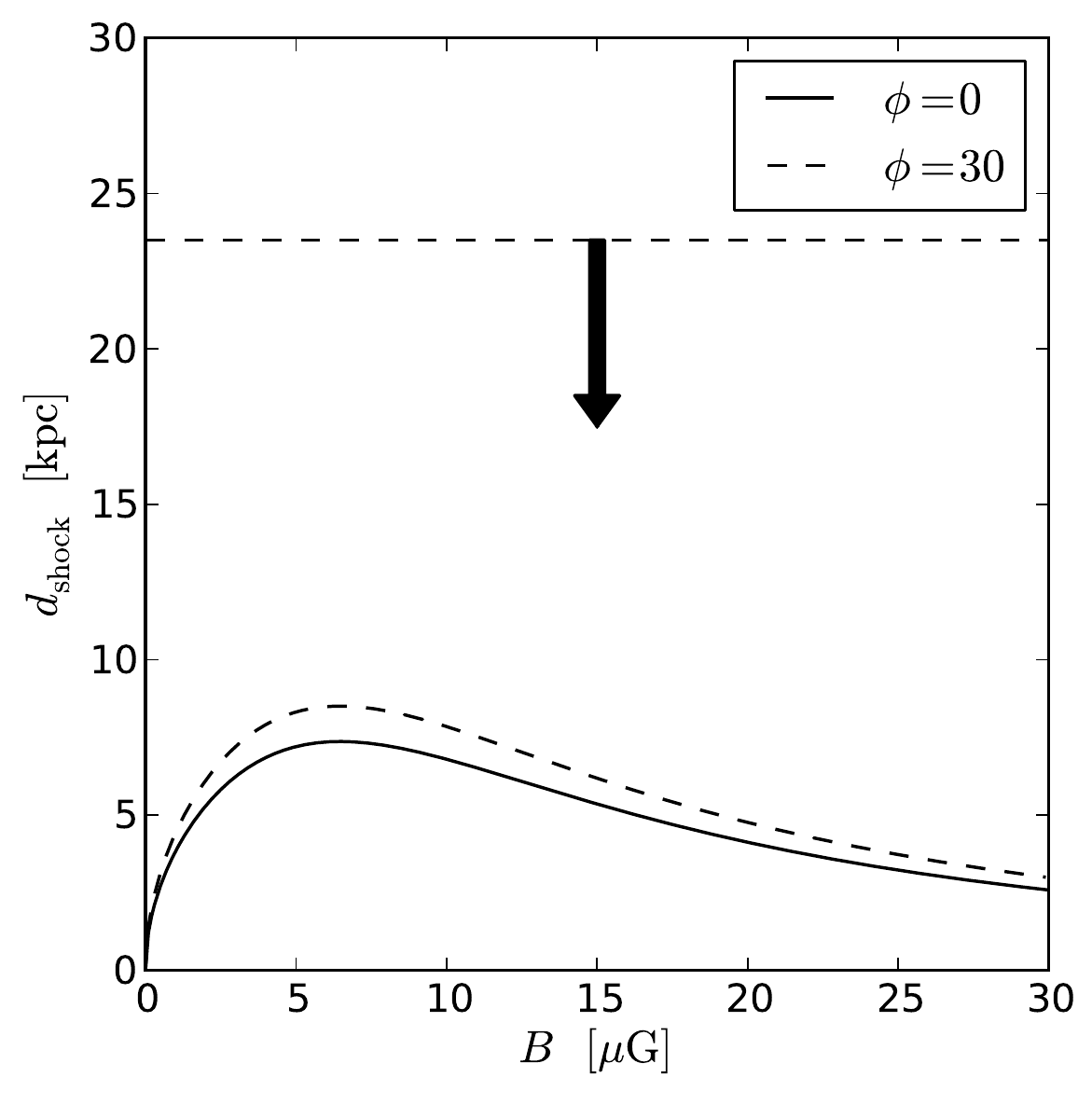}
    \caption{Predicted shock width as a function of
        magnetic field strength.  The solid and dashed
        curves show predictions of $d_{\rm shock}$ 
        using projection angles
        $\phi=0$ and $30^{\circ}$, respectively. 
        The horizontal dashed line and downward arrow
        represent the upper limit placed on the shock width
        from our $2.1\,\rm GHz$ imaging (\ShockLimit).}
    \label{f-bfield}
\end{figure}

The downstream velocity combined
with the width of the shock constrains 
$B$ via 
$d_{\rm relic} = v_{2}\times t_{\rm sync}$, where
$t_{\rm sync}$ is the characteristic timescale of
synchrotron radiation \citep[e.g., ][]{vanw11m4}:
\begin{equation}
    t_{\rm sync}= 3.2\,\times 10^{10} 
    \frac{B^{1/2}}{B^2+B_{\rm CMB}^2}
    \frac{1}{\sqrt{\nu(1+z)}}
    \,\rm yr,
    \label{e-tsync}
\end{equation}
with $\nu$ in $\rm MHz$, and $B$ and $B_{\rm CMB}$ in $\mu\rm G$.
$B_{\rm CMB}$ parametrizes electron energy loss 
by inverse Compton scattering off CMB 
photons through an equivalent
synchrotron power with magnetic field 
$B_{\rm CMB}=3.2\,{\rm \mu G} \,(1+z)^2$.  
Using $\nu=2100\,\rm MHz$, $z=0.870$, and \BCMB{}, we find 
that the predicted 
shock width is 
lower than the upper limit provided by the unresolved width
of the leading edge of the NW relic of
\ShockLimit{} (Figure \ref{f-bfield}), 
and therefore $B$ remains unconstrained using 
synchrotron timescale arguments.
Additional radio imaging with angular resolution 
$<1^{\prime\prime}$ will be required to place 
meaningful limits on $B$.

\section{Radio halo}
\label{s-halo}

\begin{figure}
    \centering
     \begin{tabular}{cc}
         \includegraphics[scale=0.55]{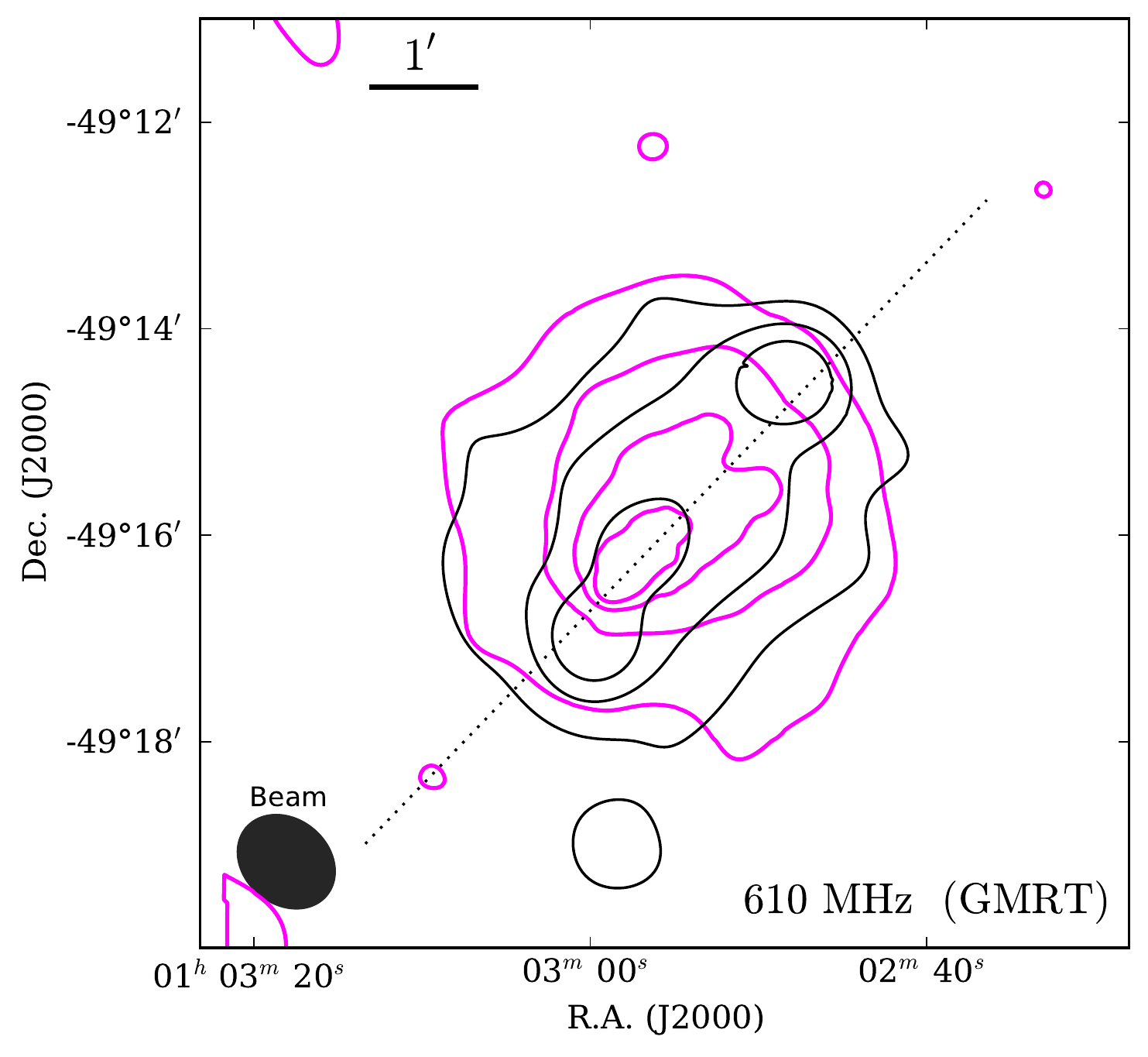} &
         \includegraphics[scale=0.55]{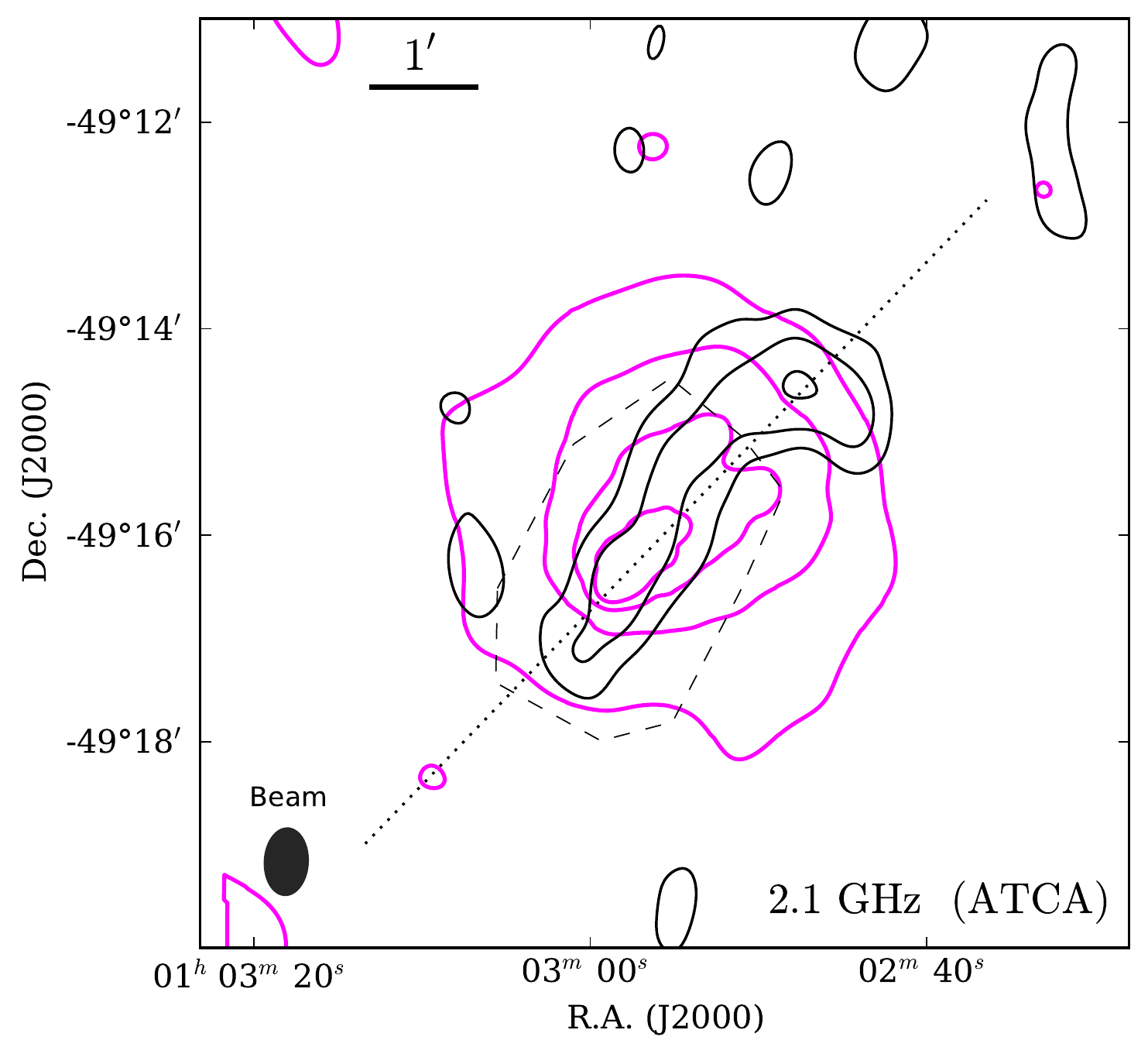}
    \end{tabular}
    \caption{El Gordo's $610\,\rm MHz$ ({\em left}) and 
             $2.1\,\rm GHz$ ({\em right}) radio halo.  Black
             contours represent the 
             surface brightess S/N with levels at $3,10,20,40,70\sigma$
             where $\sigma_{610} = 155\,\rm\mu Jy\,beam^{-1}$ and
             $\sigma_{2.1} = 28.4\,\rm\mu Jy\,beam^{-1}$.
             Purple contours represent logarithmic $0.5-2.0\,\rm keV$ 
             X-ray surface brightness \citep{mena12}.  The synthesized beam
             shapes are shown in the lower left-hand corners of the panels.
             To maximize sensitivity to the diffuse, low 
             surface-brightness halo emission, the halo images are produced 
             with $uv$ data that have had compact emission from point sources and 
             relics (i.e., most structures visible in Figures
             \ref{f-2100} and \ref{f-610}) subtracted out, and with reduced 
             spatial resolution. The dashed polygon represents the region used
             to compute the halo's integrated spectral index and excludes the area
             near the NW relic.}
    \label{f-halo}
\end{figure}

El Gordo has a powerful radio halo that
is detected at both $610\,\rm MHz$ and 
$2.1\,\rm GHz$ (Figure \ref{f-halo}), 
allowing it to join an exclusive club of 
clusters known to host both double
radio relics and halos.  Other members
include 
CL0217+70 \citep[$z=0.0655$; ][]{brow11}, 
RXCJ1314.4-2515\citep[$z=0.2439$; ][]{fere05}, 
CIZAJ2242.8+5301 \citep[$z=0.1921$; ][]{vanw10}, 
and MACS J1752.0+4440 \citep[$z=0.366$; ][]{vanweeren2012}.

\begin{figure}
    \centering
         \includegraphics[scale=0.6]{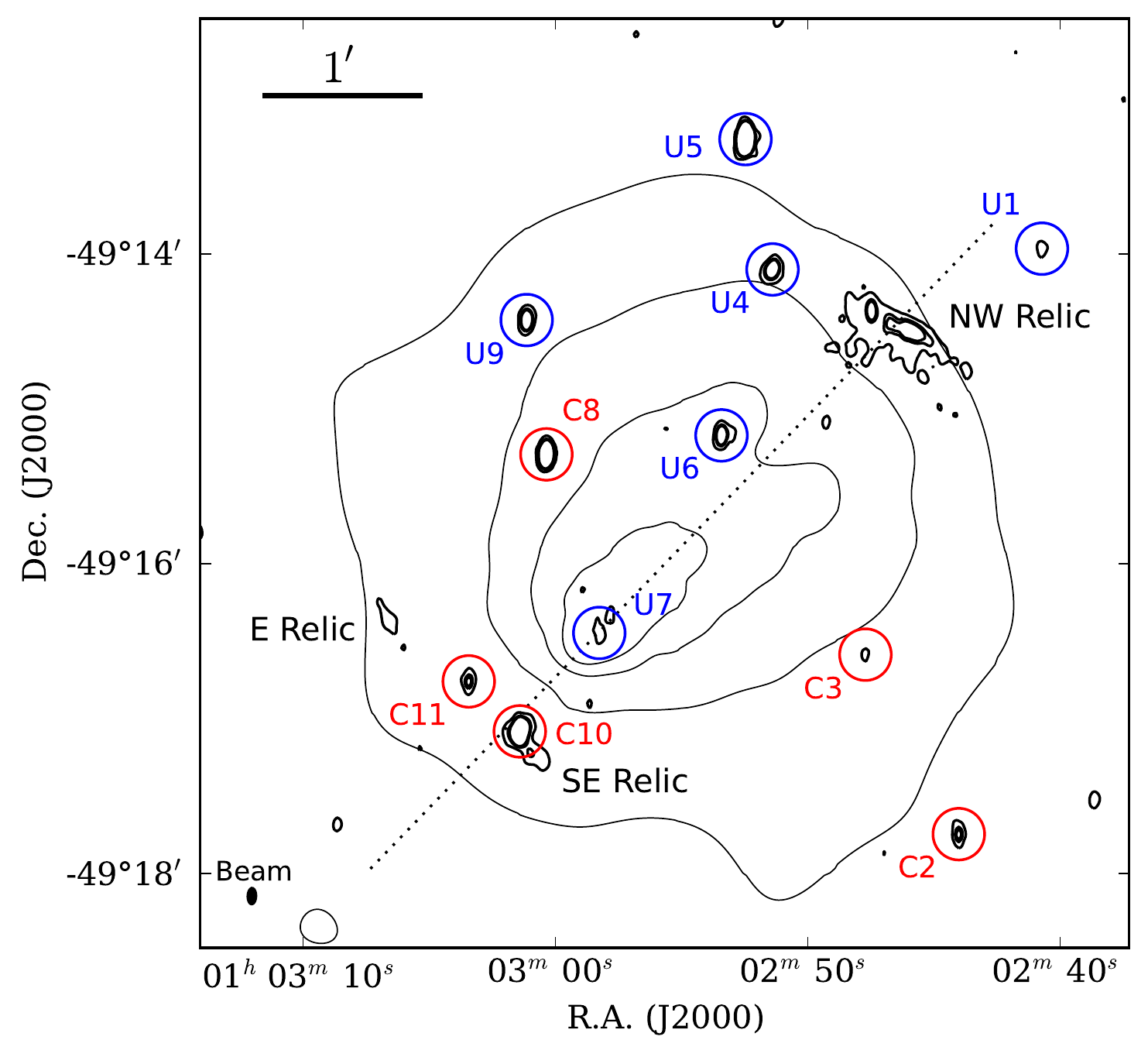}
    \caption{Locations of compact radio sources with $\rm S/N>5$ in the El Gordo
             field. Sources are labeled with prefix `C' and circled
             in red if they
             are spectroscopically-confirmed cluster 
             members \citep{sifon2013}, or labeled with prefix 'U' 
             and circled in blue if they are
             unrelated to the cluster.  Table \ref{t-sources} lists the
             sources' positions and flux densities.}
    \label{f-sources}
\end{figure}

We isolated the $610\,\rm MHz$ halo emission
by first producing an
image with \texttt{robust}$=-1$ 
from data with $uv$ distance
$d_{uv}>3.4\,\rm k\lambda$, 
corresponding to angular scales $\lesssim 1^{\prime}$, using
multi-frequency synthesis (MFS).
This image, containing only emission from compact sources,
was then Fourier transformed and subtracted from 
the $uv$ data.  The locations of compact sources with S/N$>5$
are shown in Figure \ref{f-sources}.  The point-source-subtracted $uv$ data
with $d_{uv}<3.4\,\rm k\lambda$ were then imaged using 
multi-scale clean \citep{cornwell2008} with scales
of $0^{\prime\prime}$, $30^{\prime\prime}$, and 
$90^{\prime\prime}$ and \texttt{robust}$=+1$ $uv$ 
weighting.  
The $2.1\,\rm GHz$ ATCA were imaged in the same
way, except that after subtraction of the compact emission, 
{\em all} the data were included in the final image, 
not just those with $d_{uv}<3.4\,\rm k\lambda$, 
and instead, a $uv$-taper of $30^{\prime\prime}$ was 
applied.  The ATCA data are imaged differently due to
the fact that only a small fraction have 
$d_{uv}<3.4\,\rm k \lambda$ 
(see Table \ref{t-obs-elgordo}) compared to 
the $610\,\rm MHz$ data.

The final $610\,\rm MHz$ 
and $2.1\,\rm GHz$ halo images are presented in 
Figure \ref{f-halo}.  The $610\,\rm MHz$ image 
has a sensitivity of \HaloRMS{} with a synthesized 
beam of 
\HaloBmaj{}$^{\prime\prime}\times$\HaloBmin{}$^{\prime\prime}$ 
at P.A.$=$\HaloBPA{}$^{\circ}$, and the
$2.1\,\rm GHz$ image
has a sensitivity of
\HaloAtcaRMS{} with a synthesized beam of 
\HaloAtcaBmaj{}$^{\prime\prime}\times$\HaloAtcaBmin{}$^{\prime\prime}$ 
at P.A.$=$\HaloAtcaBPA{}$^{\circ}$.
The $610\,\rm MHz$ image  recovers signal on larger spatial
scales and has a 
higher S/N than the $2.1\,\rm GHz$ image, so we use 
the $610\,\rm MHz$ map for our morphology and 
luminosity analyses, and the $2.1\,\rm GHz$ map only to
examine the halo spectral index.
Table \ref{t-properties} lists the photometric
properties of the halo.

\subsection{Geometry}

The halo is elongated in the direction of
the collision axis, as has also been observed
in the cluster merger systems
1E0657-56 \citep[Bullet cluster; ][]{mark02} and 
Abell 520 \citep[][]{gira08}.
The emission bridges the entire gap between the 
two relics lying on opposite sides of the cluster
(see Figure \ref{f-610}).  Such a complete 
emission bridge has been seen in other cluster mergers 
with radio halos \citep[e.g., ][]{bona12}.
At $610\,\rm MHz$, the halo fills
a large fraction of the projected cluster area.
If we define the effective radius of the halo 
$r_{\rm H}$ as that of a circle containing all the 
$>3\sigma$ halo emission 
(after subtracting point sources), then 
\rThreeSigmaHalo{}, which fills
$85\%$ of the projected area
within $r_{500c}$ 
\citep[$1177\pm 92\,\rm kpc$; ][]
{sifon2013}\footnote{\citet{sifon2013} report
$r_{200a}$, which we convert to $r_{500c}$ using
the conversion factor $f=1.52$ \citep{naga07}.}.
\citet{cass07} predict that halo emission in
clusters will not be self-similar, and that
the fraction of the cluster volume occupied by the
halo increases with cluster mass.  
El Gordo's halo is consistent with the
\citet{cass07} relation, which predicts
$r_{H}\sim 1.0^{+0.3}_{-0.2}\,\rm Mpc$,
and is one of the largest radio halos known.

\subsection{Spectral index}
\label{ss-spindex}

The $uv$ coverage is different
for our ATCA and GMRT datasets, and
the scale at which emission begins to be resolved
out of the ATCA data ($\sim 1.9^{\prime}$) 
is similar to the size of the halo itself.
We therefore made an unbiased comparison between
the two frequencies by first producing a multi-scale
clean model of the $610\,\rm MHz$ halo 
using data with all $uv$ distances after
subtracting the $d_{uv}>3.4\,\rm k\lambda$ point-source
image.
We then inverted the $610\,\rm MHz$ halo clean model 
and cast its $uv$ representation onto
the $uv$ coverage of the $2.1\,\rm GHz$ ATCA data.
The ``recast'' $610\,\rm MHz$ halo $uv$ data 
were then imaged identically to the $2.1\,\rm GHz$ data
(see Section \ref{s-halo}).
To remove residual relic emission from
the regions where the relics join the halo
near the ends of the cluster collision axis,
we also subtracted the $0^{\prime\prime}$-scale (point-like)
clean components from both the 
$610\,\rm MHz$ and the $2.1\,\rm GHz$ multi-scale
clean models.

\begin{figure}
    \centering
    \includegraphics[scale=0.8]{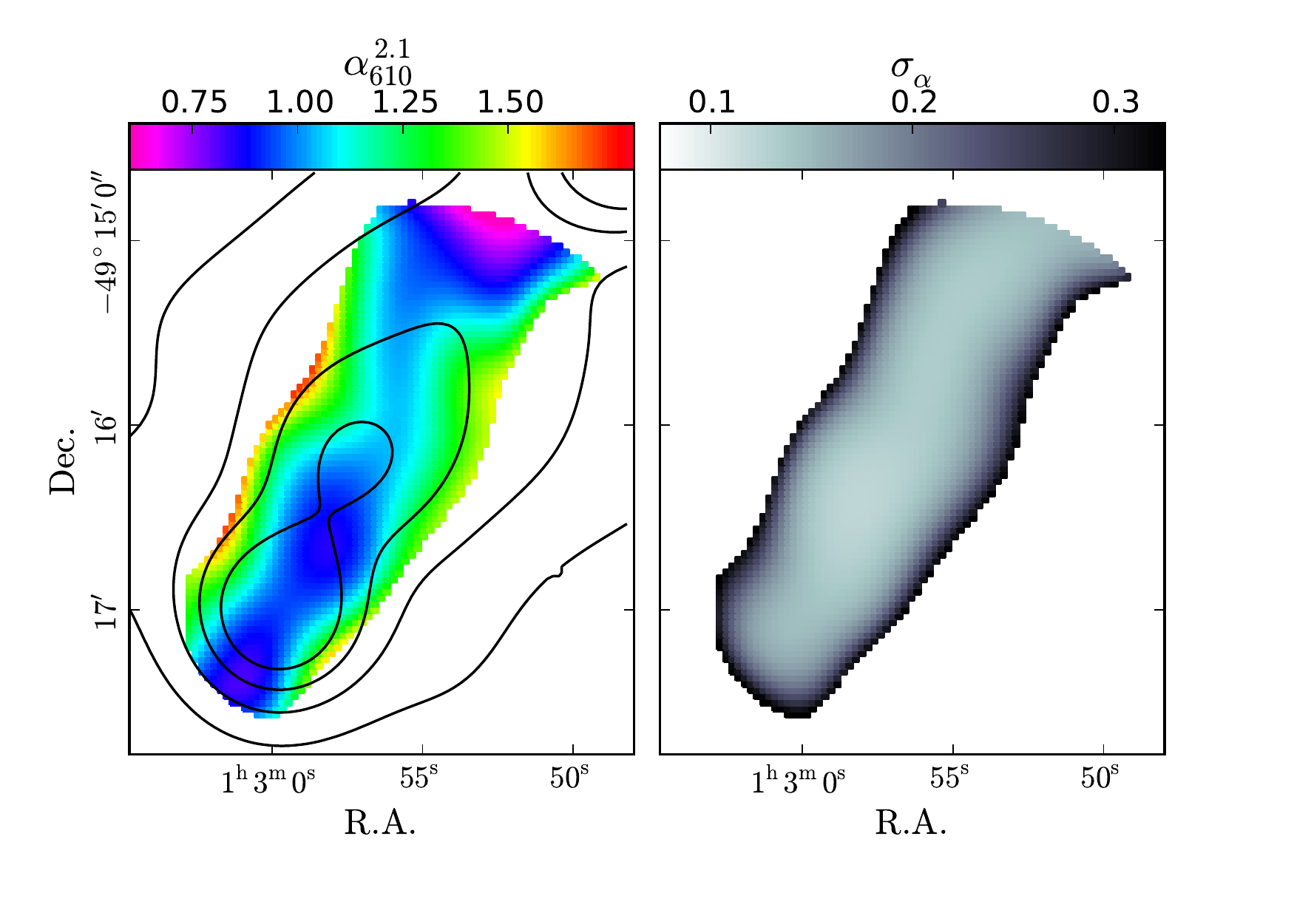}
    \caption{Spectral index $\alpha^{2.1}_{0.6}$ image of 
        El Gordo's radio halo.  
        The spectral index 
        and the uncertainty per pixel $\sigma_{\alpha}$
        are shown in the left and right panels, respectively.
        Contours represent the 
        $610\,\rm MHz$ halo intensity with levels 1, 2, 3, and 
        $4\,\rm mJy\,beam^{-1}$.}
    \label{f-index}
\end{figure}

Figure \ref{f-index} presents the spectral index
$\alpha^{2.1}_{0.6}$ image of the El Gordo halo.  
The spectral index is shallowest nearest the 
collision axis and steepens with increasing distance
from the center.  
The spectral index also flattens to the
north end of the halo where there is additional
signal from the NW relic.
The flattest
spectral index values that do not
adjoin regions containing residual relic emission
($\alpha^{2.1}_{0.6} \sim 0.75$)
 are located near the
``cold bullet'' of the merger system.
The halo's integrated spectral index is computed in a 
region that excludes the area near the NW relic 
(see Figure \ref{f-halo}) and is \haloalpha{}.
We note the importance of matching the $uv$
coverage, without which
we would have  
overestimated the integrated
spectral index to be $\alpha \sim 1.95$.
Using recent radio halo samples, \citet{fere12} 
find that clusters with $T_{\rm gas}>10\,\rm keV$ 
on average have spectral indices of $\sim 1.2$.
El Gordo is in agreement with this trend, suggesting
the halo emission is associated with
the recent energy injection caused by the ongoing
merger.  Similar merger-related spectral index
structure has been seen in 
A\,665 and A\,2163 \citep{fere04}.

\begin{figure}
    \centering
    \includegraphics[scale=0.7]{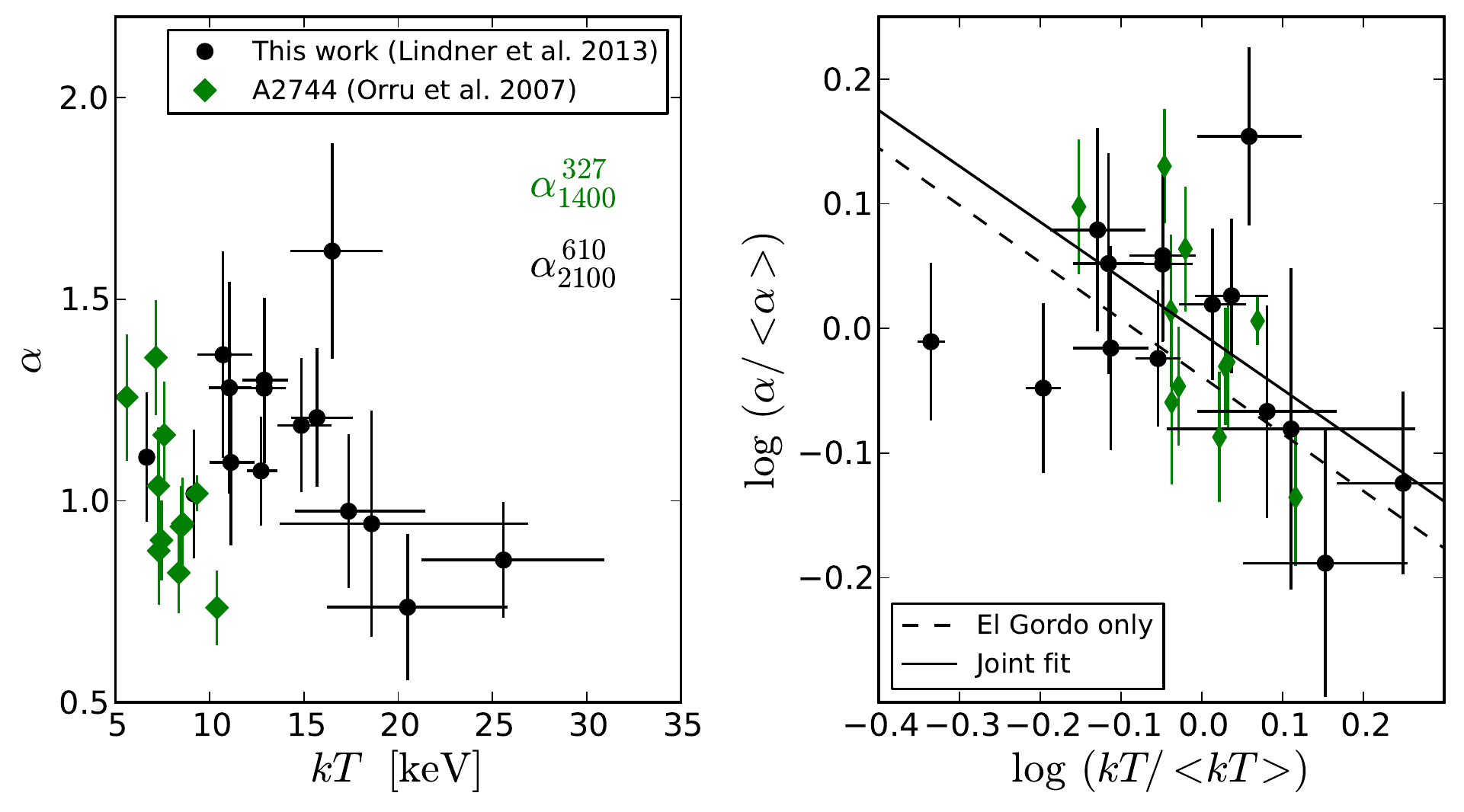}
    \caption{{\em Left:} Spatially resolved measurements of
        radio spectral index versus 
        X-ray gas temperature for  El Gordo (black points)
        compared to A2744 \citep{orru07}.  {\em Right:}
        Linear fits to the $\log$ of the data in the left panel after
        scaling by the means 
        ($\left< kT\right>_{\rm A2744}=8.0\,\rm keV$, 
        $\left< kT\right>_{\rm El\;Gordo}=14.6\,\rm keV$). 
        The dashed line shows the best-fit
        when we use El Gordo data only (power-law slope $B \simeq -0.5$), 
        and the solid line shows the best fit when we use  
        both datasets together ($B \simeq -0.4$).}
    \label{f-alpha_gas}
\end{figure}

There exists a
correlation between average
gas temperature and average 
halo spectral index in galaxy clusters with radio halos
\citep[e.g., ][]{fere12} in the sense of higher
temperature tracking flatter spectral index, which
indicates a connection
between energy injection in the ICM and halo emission.
The spatially-resolved correlation is less well
studied but remains important for understanding
systems that are not in equilibrium; these 
represent a large
fraction of halo clusters.
Recent observations have identified a resolved 
correlation between cluster gas temperature and 
halo spectral index in Abell 2744, at $z=0.31$ 
\citep[][]{orru07}.   El Gordo's spectral index
map, combined with the temperature information
derived from new $Chandra$ observations 
(Hughes et al. 2013, in prep),
allows us to characterize the spatial correlation
in a system with
gas temperatures up to $\sim 20\,\rm keV$.
Figure \ref{f-alpha_gas} 
presents a comparison of the halo
$\alpha^{2.1}_{0.6}$ versus the X-ray gas temperature
$T_{e}$ within a tiling of nearly independent  
$30^{\prime\prime}$ boxes.  
We find that the 
spectral index becomes flatter with increasing gas
temperature.  We fit a line to the 
scaled parameterized relation 
\begin{equation}
\log(\alpha/\left<\alpha \right>) =  A + B \times
\log(T_e/\left<T_e\right>),
\end{equation}
and find a best-fit power-law 
slope $B=-0.5$
using the El Gordo data alone, and $B=-0.4$
when we include the data from \citet{orru07}.
The fact that a spectral-index/gas-temperature 
correlation is found in Abell 2744 and El Gordo, systems
composed of only two merging sub-clusters 
\citep{boschin2006,mena12}, while no strong
correlation is found in 
MACS\,J0717.5+3745 or Abell\,520, systems with
$\geq 3$ merging sub-clusters 
\citep{bonafede2009,vacca2014}, suggests that
projection effects may be responsible for hiding 
the underlying correlation in clusters with 
complex merger geometries.

Radio halos produced solely by secondary electron models 
are expected to have regular morphologies
and spectral shapes that are independent
of position.  In contrast, El Gordo's halo is
aligned with the system's X-ray bullet-like ``wake'' 
and collision axis \citep{mena12}, and its 
spectral index is spatially correlated with gas
temperature.  
The fact that the radio halo in El Gordo
is associated with sites of recent energy injection 
related to the merger provides strong support for
primary models like turbulent reacceleration 
\citep{brun01} as the mechanism for production.
Recent unified models of radio halo production \citep{pfrommer2008}
explain radio halos using both primary and secondary
electron processes, with a secondary-electron dominated 
interior and primary-electron enhancements occurring 
in the low density outskirts, especially 
during mergers.  In El Gordo, the halo's wake-shaped 
morphology can be explained in this scenario,
since the secondary-electron emission is expected to trace
the cluster gas density profile (and therefore the X-ray emission), 
but the correlation between flat halo spectral indices 
and known sites of recent energy injection is indicative 
of turbulent reacceleration processes.

\subsection{Luminosity}

\begin{figure}
    \centering
    \includegraphics[scale=0.9]{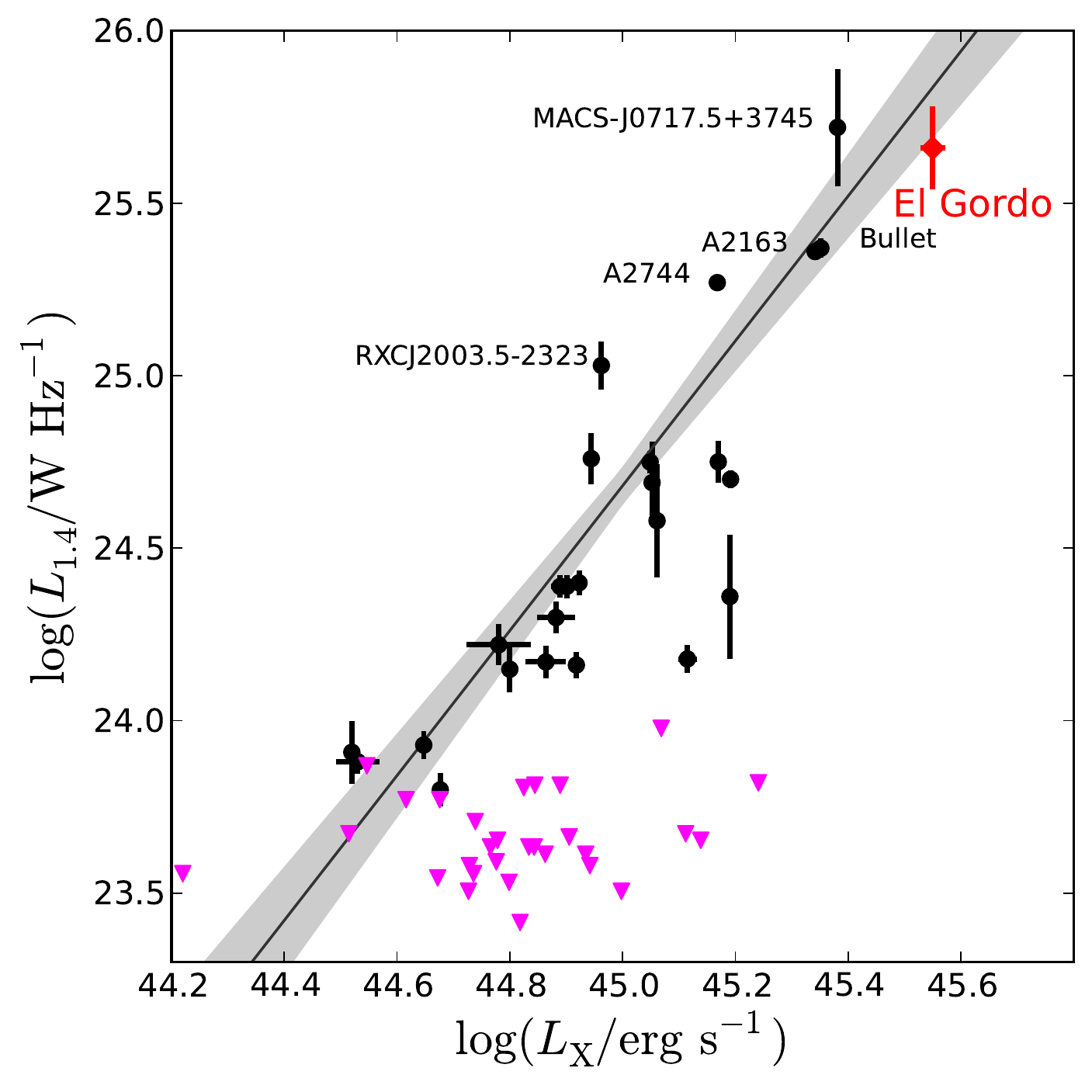}
    \caption{Rest-frame $20\,\rm cm$ spectral power
        vs. $L_{X}$ for El Gordo's and other
        halos from the literature tabulated in
        \citet{Cassano2013}.  Labels are provided for 
        halos with $\log (L_{1.4}/\rm W\,Hz^{-1})>25$.  
        The purple triangles represent upper limits.
        Errors in $\log L_{1.4}$ for El Gordo's halo include the 
        uncertainty in $\alpha^{2.1}_{0.6}$ used in the k-correction.
        The line and shaded region show the fit best correlation and 
        $\pm 1\sigma$ uncertainties from \citet{Cassano2013}.}
    \label{f-l20_lx}
\end{figure}

We compute the rest-frame $1.4\,\rm GHz$ 
spectral power 
$L_{1.4}$ of the radio halo using the 
$610\,\rm MHz$ flux density,
\Ssixten{}, which we extract  
from a $3.0^{\prime}$-radius
circle centered on the point-source-subtracted
halo image (Figure \ref{f-halo}).
After adopting the integrated spectral index \haloalpha{}
for the $k$-correction, we find \Lonepfour{}, making
El Gordo's one of the most powerful radio halos known.
Figure \ref{f-l20_lx} shows $L_{1.4}$ versus
$L_{X}$ for El Gordo compared with other clusters from
the literature. 
Only MACS\,J0717.5+3745 has greater luminosity, with
$\log (L_{1.4}/{\rm W\,Hz^{-1}})\simeq 25.70$ and spectral index 
$\alpha^{4900}_{610}=1.24\pm 0.05$ \citep{vanw09}.
Another SZE-selected cluster that hosts a radio halo
is PLCK171.9-40.7 
\citep[$\log\,(L_{1.4}/{\rm W\,Hz^{-1}})=24.70; $][]{giac13}.

\subsection{Equipartion magnetic field strength}
The magnetic field strength that minimizes
the total energy content of the halo's relativistic 
magnetized plasma occurs when the energy densities 
of the magnetic field and relativistic particles 
are approximately equal, and is known as the 
equipartition magnetic field $B_{eq}$.  $B_{eq}$
defines a natural magnetic field scale for the ICM,
and is given by
\citep{govoni2004}:
\begin{equation}
    B_{eq} = \left( \frac{24\pi}{7}u_{min}\right)^{1/2},
\end{equation}
which depends on the 
{\em k}-corrected, minimum energy density $u_{min}$ 
at redshift $z$, equal to
\begin{equation}
    \left( \frac{u_{min}}{\rm erg\,cm^{-3}} \right) = 
    \zeta(\alpha, \nu_1, \nu_2)
    (1+k)^{4/7}
    \left( \frac{\nu_{0}}{\rm MHz}\right)^{4\alpha/7}
    (1+z)^{(12+4\alpha)/7}
    \left( \frac{I_0}{\rm mJy\,arcsec^{-2}}\right)^{4/7}
    \left( \frac{d}{\rm kpc}\right)^{-4/7},
\end{equation}
in terms of a numerical factor $\zeta(\alpha, \nu_1, \nu_2)$,
spectral index $\alpha$, synchrotron spectrum integration 
limits $\nu_1$ and $\nu_2$, proton to electron 
energy density ratio $k$, observing frequency $\nu_0$,
mean halo surface brightness $I_{0}$ (\IZero{}), and 
depth $d$ ($2\times r_{\rm H}=$\depth{}). 
For comparison with $B_{eq}$ values presented for
other clusters \citep[e.g., ][]{vanweeren2009b}, 
we use $k=1$ and adopt the corresponding numerical factor
for our measured spectral index, 
$\zeta(\alpha=1.3,10\,\rm MHz,10\,GHz)=2.79\times 10^{-13}$
\citep{govoni2004},
giving \Beq{}.

We find that $B_{eq}$ is significantly greater than 
$B_{\parallel}$ ($\sim 0.01\,\rm\mu G$; Section \ref{ss-rm}).
This discrepancy exists because the two different approaches 
probe the magnetic field in very different regions of the ICM.  
$B_{eq}$ is an estimate of the mean magnetic field within the
central halo volume, approximated as a sphere with radius
\rThreeSigmaHalo{}.  On the other hand, $B_{\parallel}$
is a measure of the $n_e$-weighted parallel component of
the magnetic field in the outskirts between radii 
of $0.9\,\rm Mpc$ (the radial
position of the NW relic) to $3.6\,\rm Mpc$ (the effective radius
where the integral of the deprojected gas density converges).
Additionally, assuming the cluster is undergoing the first
pass of its collision and the NW marks the leading edge of the
shock front, the bulk of the gas at $r>0.9\,\rm Mpc$
will not yet have had its ambient magnetic field amplified 
through shock compression.

\section{Conclusions}
\label{s-conclusion}
We present new $610\,\rm MHz$
and $2.1\,\rm GHz$ observations of El Gordo, 
the highest redshift ($z=0.870$) radio halo/relic cluster
known, thereby providing important constraints 
on the non-thermal emission properties of a 
cluster merger at high redshift.

El Gordo's double radio-relic morphology 
resembles those of other massive cluster mergers
occurring in the plane of the sky, and its relic
properties are consistent with creation via
$1^{\rm st}$-order Fermi acceleration by
shocks in the ICM of the cluster collision.  
The bright leading edges of all relics remain unresolved
in our images, implying extremely thin
shock widths of \ShockLimit{}, and therefore
strong upper limits on synchrotron cooling times.
Based on relic spectral properties and previous X-ray
measurements of the gas temperature, we estimate
a shock speed of \ShockSpeed{}, and assuming
the system is undergoing the first pass of its collision, 
we constrain the angle between the collision axis and
the plane of the sky to be \collisionAngle{}.  The shallow
collision angle indicated by the system's X-ray morphology
is consistent with the high degree of integrated 
polarization  observed in the relics ($33\%$).

El Gordo's radio halo is among the largest
(\rThreeSigmaHalo{}) and most
powerful (\Lonepfour{}) known.
The halo spectral index varies with 
position, being flattest in the center along the  
collision axis and steeper away from this axis.
This spectral index morphology, 
along with the shallow integrated spectral index  
(\haloalpha{}) and the overall halo emission
shape, indicates the halo is closely related
to energy injection associated with the ongoing
merger and strongly supports primary electron
processes like turbulent reacceleration
\citep{brun01} as the halo production mechanism.

RM synthesis of the NW relic was performed using the 
wide-bandwidth polarimetry capabilities of 
the ATCA/CABB.  We find \MeanRM{} and \SigmaRM{},
and variation
between $-5$ and $+25\,\rm rad\,m^{-2}$ across the
spatially-extended relic structure.  Because there 
is little evidence for
fluctuations in $n_e$ at this position 
based on X-ray observations
\citep{mena12}, the variation is likely due 
to structure in the projected
magnetic field ($\int B_{\parallel} dl$), possibly
caused by turbulence in the ICM.
Using an estimate of the column density of
electrons along the line of sight from
X-ray observations, we estimate
typical magnetic field amplitudes of 
\Bparallel{} in the cluster outskirts.
In contrast, the volume-averaged 
equipartition magnetic field strength in the
cluster interior is \Beq{}.

The significant energy losses due to Compton
scattering off CMB photons at $z=0.870$, 
parameterized by the effective synchrotron magnetic
field strength, \BCMB{}, severely limit the radiative
lifetime of cosmic ray electrons, and new 
radio observations with angular resolution 
$<1^{\prime\prime}$ (capable
of resolving relic edges) will be required to place 
meaningful constraints on the 3D-averaged $B$ field 
using the electron
radiative lifetime.
New simulations of cluster mergers can also help us 
understand how the halo and relic properties are related 
to collision properties like the impact parameter 
and the mass ratio of the merging components.

\acknowledgments
We thank the anonymous referee whose comments have
helped to improve the quality of this manuscript.
We thank Melanie Johnston-Hollitt, Sui-Ann Mao, and
Reinout van Weeren for useful discussions.  We thank 
Phil Edwards, Robin Wark, and Shane O'Sullivan for 
help with the ATCA observations.  
We thank GMRT staff for their support during observations.
We also thank Jethro Ridl for his help in obtaining the GMRT
observations.  GMRT is an international facility run 
by the National Centre for Radio Astrophysics of 
the Tata Institute of Fundamental Research.
This research has been supported by grant NSF AST-0955810.
Partial support for this work was also provided by the 
National Aeronautics and Space Administration (NASA) 
through {\em Chandra} Award Number GO2-13156X issued to 
Rutgers University by the {\em Chandra} X-ray 
Observatory Center, which is operated by the 
Smithsonian Astrophysical Observatory for and on 
behalf of NASA under contract NAS8-03060.  

{\it Facilities:} \facility{ATCA}, 
                  \facility{GMRT}, 
                  \facility{{\em Chandra X-ray Observatory}}

\clearpage

\clearpage

\clearpage

\clearpage

\begin{deluxetable}{lccccc}
    \tablecaption{Radio observations \label{t-obs-elgordo}}
    \tablewidth{0pt}
    \tablehead{
        \colhead{} & \colhead{} & 
        \colhead{} & \colhead{} &
        \colhead{$t_{\rm obs}$} &
        \colhead{$B_{\rm min}$--$B_{\rm max}$}
        \\
        \colhead{Telescope} &  \colhead{Frequency} & 
        \colhead{Configuration} & \colhead{Obs date} &
        \colhead{(hr)} &
        \colhead{($\rm k\lambda$)} 
              }
    \startdata
        ATCA      &  $2.1\,\rm GHz$  &  6A     &  Dec 2011    &  12 & 2.4--41.6 \\
                  &                  &  1.5B   &  April 2012  &  8  & 1.4--30.1 \\[0.5ex]
        GMRT      &  $610\,\rm MHz$  &  fixed  &  Aug 2012    &  12 & 0.48--52.8    
    \enddata
\end{deluxetable}

\begin{deluxetable}{cllcccc}
    \tablecaption{Radio sources}
    \tablewidth{0pt}
    \tablehead{
        \colhead{} & 
        \colhead{R.A.\tablenotemark{a}} & 
        \colhead{Dec.\tablenotemark{a}} & 
        \colhead{$S_{610}\tablenotemark{b}$} & 
        \colhead{$S_{2100}\tablenotemark{b}$} & \colhead{}
        \\
        \colhead{ID} & 
        \colhead{(h:m:s)} & 
        \colhead{($\circ:\prime:\prime\prime$)} & 
        \colhead{($\mu$Jy)} & 
        \colhead{($\mu$Jy)} 
        }
        \startdata
            U1 & 01:02:40.74  & -49:13:58.09  & -             & $ 104\pm 15$ \\
C2 & 01:02:44.02  & -49:17:44.90  &  $ 614\pm 77$ & $ 253\pm 19$ \\
C3 & 01:02:47.72  & -49:16:35.41  &  $ 235\pm 53$ & $  87\pm 17$ \\
U4 & 01:02:51.40  & -49:14:06.21  &  $1953\pm 205$ & $ 606\pm 34$ \\
U5 & 01:02:52.47  & -49:13:15.71  &  $5085\pm 511$ & $5044\pm 253$ \\
U6 & 01:02:53.42  & -49:15:10.50  &  $1021\pm 112$ & $ 503\pm 29$ \\
U7 & 01:02:58.26\tablenotemark{c} & -49:16:27.01\tablenotemark{c} & $ 590\pm 96$ & -              \\
C8 & 01:03:00.35  & -49:15:17.83  &  $1761\pm 182$ & $1612\pm 82$ \\
U9 & 01:03:01.13  & -49:14:25.79  &  $1068\pm 118$ & $ 549\pm 31$ \\
C10 & 01:03:01.41 & -49:17:05.11  &  $5750\pm 579$ & $1859\pm 94$ \\
C11 & 01:03:03.43 & -49:16:45.73  &  $ 331\pm 50$  & $ 247\pm 19$ 

        \enddata
        \tablenotetext{a}{Position of $2.1\,\rm GHz$ counterpart}
        \tablenotetext{b}{Total flux densities extracted using the AIPS task \texttt{sad}}
        \tablenotetext{c}{Position of $610\,\rm MHz$ counterpart}
        \label{t-sources}
\end{deluxetable}

\begin{deluxetable}{ccccccccc}
    \tablecaption{Relic and halo properties}
    \tablewidth{0pt}
    \tablehead{
        \colhead{  } & 
        \colhead{ R.A. } & 
        \colhead{ Dec. } & 
        \colhead{ $S_{610}$ } & 
        \colhead{ $S_{843}\tablenotemark{a}$ } & 
        \colhead{ $S_{2100}$ } & 
        \colhead{  } & 
        \colhead{  } &
        \colhead{ $L_{1.4}$ }
        \\
        \colhead{Name} & 
        \colhead{ (h:m:s) } & 
        \colhead{ ($\circ:\prime:\prime\prime$)} & 
        \colhead{ (mJy) } & 
        \colhead{ (mJy) } & 
        \colhead{ (mJy) } & 
        \colhead{$\alpha_{0.6}^{2.1}$} & 
        \colhead{$\alpha_{1.6}^{2.6}$} &
        \colhead{ $(\log[\rm W\,Hz^{-1}])$ }
            }
    \startdata
    NW Relic   & 01:02:46 & -49:14:43 & $19 \pm 2$   & $18.2$ & $4.3 \pm  0.2$ & $1.19\pm 0.09$ & $2.0\pm 0.2$ & $25.49\pm 0.11$ \\
E Relic    & 01:03:07 & -49:16:16 & $1.2\pm 0.2$ & -      & $0.41\pm 0.04$ & $0.9 \pm 0.1$  & $1.2\pm 0.5$ & $24.32\pm 0.17$ \\
SE Relic   & 01:03:01 & -49:17:14 & $3.0\pm 0.3$ & -      & $0.48\pm 0.04$ & $1.4 \pm 0.1$  & $1.2\pm 0.4$ & $24.65\pm 0.13$ \\
Halo       & 01:02:55 & -49:15:37 & $29 \pm 3$   & -      & $2.43\pm 0.18$ & $1.2 \pm 0.1$  & -            & $25.66\pm 0.12$     

    \enddata
    \tablenotetext{a}{Flux density from SUMSS \citep{mauc03}}
    \label{t-properties}
\end{deluxetable}

\end{document}